\documentclass[reprint,amsmath,amssymb,aps,prd,nofootinbib,floatfix,superscriptaddress,preprintnumbers]{revtex4-2}

\usepackage{subcaption}
\usepackage{multirow}
\usepackage{graphicx}
\usepackage{bm}
\usepackage{xcolor}
\usepackage{mathrsfs}
\usepackage{amssymb}
\usepackage{hyperref}
\usepackage[normalem]{ulem}
\usepackage{amsmath}
\allowdisplaybreaks

\usepackage{tikz}
\usetikzlibrary{decorations.pathmorphing,decorations.markings,calc}

\hypersetup{colorlinks=true,linkcolor=blue,citecolor=blue,urlcolor=blue}
\usepackage{tikz}
\usetikzlibrary{decorations.pathmorphing,decorations.markings}
\tikzset{
  grav/.style={decorate,decoration={snake,amplitude=1.0pt,segment length=3.6pt,pre length=1pt,post length=1pt},line width=0.45pt},
  wline/.style={dotted,line width=1.1pt},
  qprop/.style={double,double distance=1.4pt,line width=0.4pt,postaction={decorate},
     decoration={markings,mark=at position 0.6 with {\arrow[line width=0.5pt]{>}}}},
  qdot/.style={circle,fill=black,inner sep=0pt,minimum size=4pt},
  mdot/.style={circle,draw=black,fill=white,inner sep=0pt,minimum size=4.2pt,line width=0.55pt},
  mcross/.style={circle,draw=black,fill=white,inner sep=0pt,minimum size=6pt,line width=0.55pt,
     path picture={\draw[line width=0.5pt] (path picture bounding box.south west) -- (path picture bounding box.north east)
     (path picture bounding box.north west) -- (path picture bounding box.south east);}},
  vlab/.style={font=\scriptsize,inner sep=1pt},
}

\newcommand*\diff{\mathop{}\!\mathrm{d}}
\newcommand*\Diff{\mathop{}\!\mathrm{D}}
\tikzset{
  grav/.style={decorate,decoration={snake,amplitude=1.05pt,segment length=4pt,pre length=1pt,post length=1pt},line width=0.45pt},
  wline/.style={dotted,line width=1.1pt},
  qprop/.style={double,double distance=1.5pt,line width=0.4pt,postaction={decorate},
     decoration={markings,mark=at position 0.6 with {\arrow[line width=0.5pt]{>}}}},
  qdot/.style={circle,fill=black,inner sep=0pt,minimum size=4.4pt},
  mdot/.style={circle,draw=black,fill=white,inner sep=0pt,minimum size=4.6pt,line width=0.55pt},
  mcross/.style={circle,draw=black,fill=white,inner sep=0pt,minimum size=6.4pt,line width=0.55pt,
     path picture={\draw[line width=0.5pt] (path picture bounding box.south west) -- (path picture bounding box.north east)
     (path picture bounding box.north west) -- (path picture bounding box.south east);}},
  blob/.style={circle,draw=black,fill=black!12,inner sep=0pt,minimum size=15pt,line width=0.55pt,font=\footnotesize},
  vlab/.style={font=\footnotesize,inner sep=1.5pt},
}

\DeclareCaptionJustification{fulljust}{\leftskip=0pt\rightskip=0pt\parfillskip=0pt plus 1fil\relax}
\begin{document}

\preprint{HU-EP-26/34-RTG}

\title{Renormalized dynamical tidal response of neutron stars in binary dynamics}

\author{M. V. S. Saketh}
\email{venkata.saketh@icts.res.in}
\affiliation{International Centre for Theoretical Sciences, Tata Institute of Fundamental Research, Bangalore 560089, India}
\author{Raj Patil}
\email{raj.patil@aei.mpg.de}
\affiliation{Max Planck Institute for Gravitational Physics (Albert Einstein Institute), Am M\"uhlenberg 1, D-14476 Potsdam, Germany}
\affiliation{Institut f{\"u}r Physik und IRIS Adlershof, Humboldt-Universit{\"a}t zu Berlin, Zum Großen Windkanal 2, D-12489 Berlin, Germany}
\affiliation{Center of Gravity, Niels Bohr Institute, Blegdamsvej 17, 2100 Copenhagen, Denmark}
\author{Godwin Martin}
\email{godwin.martin@icts.res.in}
\affiliation{International Centre for Theoretical Sciences, Tata Institute of Fundamental Research, Bangalore 560089, India}
\author{Nils Andersson}
\email{N.A.Andersson@southampton.ac.uk}
\affiliation{Mathematical Sciences and STAG Research Centre, University of Southampton, Southampton, United Kingdom}
\author{Suprovo Ghosh}
\email{S.Ghosh@soton.ac.uk}
\affiliation{Mathematical Sciences and STAG Research Centre, University of Southampton, Southampton, United Kingdom}
\author{Jan Steinhoff~ }
\email{jan.steinhoff@aei.mpg.de}
\affiliation{Max Planck Institute for Gravitational Physics (Albert Einstein Institute), Am M\"uhlenberg 1, D-14476 Potsdam, Germany}
\affiliation{Institut f{\"u}r Physik und IRIS Adlershof, Humboldt-Universit{\"a}t zu Berlin, Zum Großen Windkanal 2, D-12489 Berlin, Germany}

\date{\today}

\begin{abstract}

We present the relativistic dynamical tidal response of a neutron star computed using stellar perturbation theory and matched to a worldline effective field theory via the tidal scattering phase. This phase is defined as the difference between the Compton/Raman
scattering phase of gravitational waves from a neutron star and the real part of the black-hole scattering phase.
The gravitational corrections to the response function are obtained using self-energy corrections up to fourth post-Minkowskian order.
The response function for neutron stars obeying the BSk22 equation-of-state from the Brussels-Skyrme family is then consistently matched to an effective theory of relativistic quadrupolar harmonic oscillators to model the fundamental mode  ($f$-mode) and low-frequency gravity modes ($g$-modes).
This matching includes the frequency and tidal coupling of each mode and the post-adiabatic dynamical tidal Love number, and once matched this model is in excellent agreement with the full neutron star response.
With the external tidal field up to third post-Newtonian order, we analyse the tidal coupling of the induced quadrupole on the neutron star in a circular binary system and work out the adiabatic binding energy.
\end{abstract}

\maketitle
\section{Introduction}
\label{sec-intro}

Coalescence of binaries containing compact objects is one of the primary sources of detectable gravitational radiation. The compact objects are likely black holes and neutron stars. For the latter, gravitational-wave (GW) detection has the potential to be a powerful probe into the nature of matter under extreme conditions. This requires identifying the imprint of physics in the stellar interior onto GWs. Tidal effects constitute an important such imprint. They arise from the induced deformation of neutron stars due to their companion's gravitational field, and the deformation is determined by the equation of state and the matter composition in the stellar interior.

The tidal deformability may be roughly split into two main components. The \emph{static tide}, represented by the Love numbers~\cite{Hinderer:2007mb,Rodriguez:2026iot,Chakraborty:2026qru}, was already constrained by the GW170817 data~\cite{LIGOScientific:2018cki}. The static tidal response, similar to bulk stellar properties such as the mass and radius, is primarily determined by the equilibrium equation of state (EoS). In contrast, the \emph{dynamical tide} probes the star's oscillation modes~\cite{Lai:1993di}, which are more sensitive to its internal composition and microphysics, thereby connecting tidal dynamics with neutron-star asteroseismology. Detection of these features will likely require the sensitivity of the next generation of detectors, such as Cosmic Explorer~\cite{Reitze:2019iox} and the Einstein Telescope~\cite{ET:2025xjr}, planned for the next decade.
The dominant dynamical tide contribution associated with the star's fundamental oscillation mode ($f$-mode) has already been included in state-of-the-art waveform models~\cite{Hinderer2016EffectsON,fmode_pratten, NRttidalv3} used in the analysis of signals~\cite{Pratten:2020fqn,Wen_2019,NRttidalv3}. However, these models neglect various low-frequency resonances which depend directly on the composition and state of matter~\cite{Counsell_2025,Yu_2017, Gittins:2026ntx,Tran:2022dva,Thakur:2607.20693}, and assume that the dynamical response can be understood as that of a collection of harmonic oscillators. The latter has only been rigorously demonstrated in the Newtonian case and is not yet shown for relativistic stars, although relativistic forced-oscillator equations and overlap integrals have been derived~\cite{HegadeKR:2025qwj}, and an oscillator model of the dominant $f$-mode remains a good approximation in the relativistic case~\cite{HegadeKR:2026kku}. Even if the oscillator description remains valid in the relativistic regime, a consistent prescription is required to determine its parameters, in particular the mode amplitudes and frequencies, from the stellar response.

The last two decades have witnessed the rise of effective field theories (EFT)~\cite{Goldberger:2004jt,Levi:2018nxp,Porto:2016pyg} applied to the two-body problem in general relativity, with remarkable progress \cite{Foffa:2019hrb,Blumlein:2019zku,Blumlein:2020pyo,Blumlein:2021txe,Almeida:2021xwn,Porto:2024cwd,Porto:2026fsd,Amalberti:2024jaa,Brunello:2025gpf,Brunello:2026anu,Brunello:2026mkv,Blumlein:2021txj,Almeida:2026clf} including the incorporation of spin~\cite{Porto:2005ac,Levi:2015msa,Kim:2022pou,Mandal:2022nty,Kim:2022bwv,Mandal:2022ufb}, and also finite-size effects in a systematic post-Newtonian (PN) expansion (for other works, see Refs.~\cite{lorentzanddroste,Einstein:1938yz,
Futamase:2007zz,Blanchet:2013haa,Schafer:2018kuf,Jaranowski:1997ky,
Damour:2014jta,Jaranowski:2015lha,Bernard:2015njp,Bernard:2016wrg,Damour:2016abl,
Blanchet:2023sbv,Blanchet:2023bwj}).
Here, the compact object is replaced by a worldline action, which at leading PN order is just that of a test particle, but includes additional degrees of freedom and/or non-minimal couplings to the tidal field at higher orders.
For the tidal effects, the effective Hamiltonian for conservative dynamical gravitoelectric tides was first computed in Refs.~\cite{Vines:2010ca,Steinhoff:2016rfi} up to 1PN and then extended up to 2PN in Ref.~\cite{Mandal:2023lgy}. The state-of-the-art of the conservative dynamical gravitoelectric tides is the 3PN effective Hamiltonian computed in Ref.~\cite{Mandal:2023hqa}. In the adiabatic limit, the 2PN effective Hamiltonian was computed in Refs.~\cite{Bini:2012gu,Henry:2019xhg} for both gravitoelectric and gravitomagnetic tides. Also in this case, the state-of-the-art is the 3PN effective Hamiltonian computed in Ref.~\cite{Mandal:2023hqa}.
For the radiative dynamics, the NLO (1PN) was computed by \cite{Vines:2011ud,Banihashemi:2018xfb,Landry:2018bil}, whereas the NNLO (2PN) corrections were computed in~\cite{Mandal:2024iug,PhysRevD.108.089901}. Here the state-of-the-art is the 2.5PN result computed in \cite{Dones:2024odv}.
The EFT tidal response is essentially the coefficients or functions tuning the non-minimal couplings and the dynamics of the tidal degrees of freedom. Once the tidal response is determined, there are well-established methods to construct binary Hamiltonians~\cite{Mandal:2023hqa}. However, determining the EFT tidal response is a very difficult problem~\cite{Jarequi:2026cyp,Apostolidis:2026qsg,Chakrabarti:2013lua,Gao:2026inprep}. In parallel, the frequency-dependent response has been computed directly from relativistic stellar perturbations by matching to an exterior tidal solution~\cite{HegadeKR:2024agt,Andersson:2025iyd}. These calculations resolve the oscillation modes, but determining the tidal response within an EFT framework while including resonances remains desirable. As mentioned above, binary Hamiltonians require the response as the coefficients of the worldline couplings.

A procedure to compute the tidal response in a post-Minkowskian (PM)
\cite{Westpfahl:1979gu,Westpfahl:1980mk,Damour:2016gwp,Buonanno:2022pgc,Travaglini:2022uwo,Bjerrum-Bohr:2022blt,Kosower:2022yvp,Bern:2025wyd, Driesse:2026qiz}
expansion from stellar/black-hole perturbation theory \cite{Thorne1967,Lindblom:1983ps,Detweiler1985,Chandrasekhar:1991fi,PhysRev.108.1063,Zerilli:1970se,Teukolsky:1972my,Teukolsky:1973ha,Mano:1996vt,Mano:1996mf,Mano:1996gn,Sasaki:2003xr,Dolan:2007ut,Dolan:2008kf,Casals:2015nja} by matching the gravitational Compton/Raman scattering amplitudes \cite{Arkani-Hamed:2017jhn,Guevara:2018wpp,Chung:2018kqs,Aoude:2020onz,Chiodaroli:2021eug,Bautista:2021wfy,Aoude:2022trd,Bjerrum-Bohr:2023iey,Cangemi:2023bpe,Ben-Shahar:2023djm,Scheopner:2023rzp,Bautista:2023sdf,Bautista:2024emt,Haddad:2024ebn,Vazquez-Holm:2025ztz,Correia:2024jgr,Akpinar:2025huz,Akpinar:2025byi,Ivanov:2022qqt,Saketh:2023bul,Saketh:2024juq,Ivanov:2026icp,Bjerrum-Bohr:2025bqg,Bautista:2026qse,Bjerrum-Bohr:2026fhs,Bautista:2026g4,Brunello:2026c5,Correia:2026utp,Bjerrum-Bohr:2026low} in the EFT with those in perturbation theory was introduced in Ref.~\cite{Saketh:2026trm}. The tidal-response function was explicitly determined at leading and partially at sub-leading PM order, and it was demonstrated that the dynamical tidal response thus obtained numerically for two stellar models is consistent with the static Love number in the low-frequency limit, while also capturing the relevant $f$- and $g$-mode signatures, including the finite width of the former owing to gravitational radiation. Since this approach matches a gauge-invariant scattering observable, it provides a convenient way to constrain the EFT tidal-response function. In this work, we present the computation to relative 4PM order, which determines the tidal-response function to the accuracy needed for relative 6PN tidal effects in binary dynamics and waveforms. The 4PM tidal-response function, obtained by matching scattering observables, resolves the $f$- and $g$-mode resonances of the isolated star. We also discuss the renormalization of the tidal-response function. We then match it to an EFT of relativistic quadrupolar harmonic oscillators for the $f$- and $g$-modes, fixing the frequency and tidal coupling of each mode and the post-adiabatic Love number, and find excellent agreement with the full response. Finally, we show how the tidal response may be incorporated into the binary Hamiltonian obtained previously in Ref.~\cite{Mandal:2023hqa}. This is the first consistent inclusion of a realistic, dynamical tidal-response function into the 3PN dynamics. However, computing observables from the Hamiltonian close to the resonant frequencies also requires a careful treatment of radiative effects, a step that remains to be completed. For now, we compute the conservative 3PN coupling energy of the induced quadrupole excluding a small neighbourhood around resonant crossings, and the conservative 3PN tidal binding energy in the post-adiabatic limit. We also present results that probe the relevance of higher-order corrections to the tidal response, and the deviation from the oscillator model currently used to capture tidal effects.

The paper is organized as follows. In Section~\ref{sec-recap}, we review the worldline EFT and the definition of the tidal phase, and the method established in Ref.~\cite{Saketh:2026trm}. In Section~\ref{sec-resum}, we compute the self-energy of the quadrupole propagator as a Dyson resummation through $O(M^4)$, and discuss the renormalization of divergences. In Sec.~\ref{sec:ntider},  we derive the renormalized tidal response, discuss the choice of scale, and study the response for a $1.4\,M_\odot$ star using the BSk22 EoS model~\cite{2013PhRvC..88f1302G,2018MNRAS.481.2994P}. In Sec.~\ref{sec-osc}, we match the numerically obtained response to an action of relativistic harmonic oscillators for $1.4\,M_\odot$ and $1.98\,M_\odot$ stars, with the fitted parameters of both given in Table~\ref{tab:osc} and the fit plotted for the $1.4\,M_\odot$ star. In Sec.~\ref{sec-binary}, we incorporate the tidal response into binary dynamics at 3PN, through the tidal energy of the induced quadrupole and the adiabatic binding energy, for equal-mass binaries. In Sec.~\ref{sec-conclusion}, we conclude with a short discussion of the work, its scope and future directions.

We use the metric signature $(-,+,+,+)$ and units $G=c=1$.
The mass $M$ stands for $GM/c^2$. The Planck mass $m_{\rm pl}$ tracks powers of $G$ instead. We use dimensional regularization for divergent integrals, and thus define $\mu$ as the renormalization scale, and $d=4-2\epsilon$ as the number of spacetime dimensions\footnote{The $\epsilon$ in Ref.~\cite{Mandal:2023hqa} is related to the epsilon defined here as $\epsilon_{\rm ref}=-2\epsilon$.}. We have $m_{\rm pl}=\tilde{\mu}^{\epsilon}(32\pi)^{-\frac{1}{2}}=\left(\frac{\mu^2}{4\pi e^{\gamma_{\rm E}}}\right)^{\epsilon/2}(32\pi)^{-\frac{1}{2}}$. The PM expansion parameter is $\bar{\omega}=M\omega/(32\pi m_{\rm pl}^2)=\tilde{\mu}^{-2\epsilon}M\omega$. We use the $\overline{\rm MS}$~\cite{Peskin:1995ev} prescription when defining counterterms to cancel UV divergences.

\section{Isolating tidal effects in EFT and perturbation theory}
\label{sec-recap}
In this section, we review the techniques developed in Ref.~\cite{Saketh:2026trm}. The key idea is that in a worldline EFT, the classical part of the S-matrix, $S=e^{2i\Delta}$, for gravitational waves scattering off a conservative compact object can be factorized as
\begin{equation}
    e^{2i\Delta} = 1+iT_{\rm pp} +iT_{\rm tid} = (1+iT_{\rm tid}+T_{\rm tid}T_{\rm pp}^\dagger)(1+i T_{\rm pp}) \ .
\end{equation}
Here, the second equality is an identity that follows from the unitarity of the point-particle S-matrix, $(1+iT_{\rm pp})^\dagger(1+iT_{\rm pp})=1$, and it splits the full S-matrix into a point-particle factor and a factor that contains all tidal effects.
The $T_{\rm tid}$ is the transition operator associated with tidal effects (including their gravitational dressing), while $T_{\rm pp}$ is the transition operator associated with the point particle, not including tidal effects.

The worldline action including quadrupolar tidal effects can be written as
\begin{equation}
\begin{aligned}
\mathcal{S}={}&\mathcal{S}_{\rm EH}+\mathcal{S}_{\rm GF}-M\int \diff \tau+\mathcal{S}_Q-\frac12\int \diff\tau\,Q^{\mu\nu}E_{\mu\nu}\ ,
\end{aligned}
\end{equation}
where $\mathcal{S}_{\rm EH}$ and $\mathcal{S}_{\rm GF}$ are the Einstein-Hilbert and the corresponding gravitational gauge-fixing actions, respectively. The remaining terms describe the point particle of mass $M$, its electric-quadrupolar dynamics, and the coupling of its quadrupole to the electric part of the Weyl tensor $C_{\mu \alpha \nu \beta}$, defined as $E_{\mu\nu}={}-C_{\mu\alpha\nu\beta}\,u^\alpha u^\beta$, where $u^{\alpha}$ is the four-velocity of the point particle.
The conservative tidal response is then captured via the propagator\footnote{Note the difference in sign conventions compared to Ref.~\cite{Saketh:2026trm}. The tidal response $F$ is defined here with the opposite sign compared to Ref.~\cite{Saketh:2026trm}, so that there is no relative minus sign between $F$ and $k_2$. Additionally, the tidal coupling and the relativistic tidal field have the opposite signs.}
\begin{equation}
\begin{gathered}
\big\langle \mathbb{T}\tilde Q^{\mu\nu}(\omega)\tilde Q^{\rho\sigma}(\omega')\big\rangle=-iR^5F_{\rm B}(\omega)\,\Pi_d^{\mu\nu,\rho\sigma}\,2\pi\delta(\omega+\omega')\ ,\\
\end{gathered}
\label{eq:Qprop}
\end{equation}
where $\mathbb{T}$ denotes the time-ordering symbol, and $\tilde Q^{\mu\nu}(\omega)$ is the Fourier-transform of the quadrupole. The symbol $\Pi_d^{\mu\nu,\rho\sigma}$ denotes a projector defined as
\begin{equation}
\begin{split}
    \Pi_d^{\mu\nu,\rho\sigma}&=\tfrac12\big(\mathcal P^{\mu\rho}\mathcal P^{\nu\sigma}+\mathcal P^{\mu\sigma}\mathcal P^{\nu\rho}\big)-\tfrac{1}{d-1}\mathcal P^{\mu\nu}\mathcal P^{\rho\sigma},\\
\mathcal P^{\mu\nu}&=\eta^{\mu\nu}+u^\mu u^\nu  \ .
\end{split}
\end{equation}
In $d$-spacetime dimensions, $\mathcal {P}^{\mu\nu}$ projects orthogonally to the four-velocity $u^{\mu}$, while $\Pi_d^{\mu\nu,\rho\sigma}$ projects onto symmetric trace-free (STF) tensors in this subspace.

We can diagrammatically represent the transition operator for a graviton scattering off the worldline in two parts. The first part, not involving quadrupolar tides, takes the form
\begin{equation}
\begin{split}
iT_{\rm pp}=
\begin{tikzpicture}[baseline=0.35cm]
  \draw[grav] (-0.7,0.95) -- (0,0);
  \draw[grav] (0,0) -- (0.7,0.95);
  \draw[wline] (-0.75,0) -- (0.75,0);
  \node[blob] at (0,0) {$M$};
\end{tikzpicture}
\;&=\;
\begin{tikzpicture}[baseline=0.35cm]
  \draw[grav] (-0.75,0.95) -- (0,0.55);
  \draw[grav] (0,0.55) -- (0.75,0.95);
  \draw[grav] (0,0) -- (0,0.55);
  \draw[wline] (-0.75,0) -- (0.75,0);
  \node[mdot,label={[vlab]below:$M$}] at (0,0) {};
\end{tikzpicture}
\;+\;
\begin{tikzpicture}[baseline=0.35cm]
  \draw[grav] (-0.7,0.95) -- (0,0);
  \draw[grav] (0,0) -- (0.7,0.95);
  \draw[wline] (-0.75,0) -- (0.75,0);
  \node[mcross,label={[vlab]below:$M$}] at (0,0) {};
\end{tikzpicture}\\
&\quad+\;
\begin{tikzpicture}[baseline=0.35cm]
  \draw[grav] (-1.0,0.95) -- (-0.4,0.55);
  \draw[grav] (-0.4,0.55) -- (0.4,0.55);
  \draw[grav] (0.4,0.55) -- (1.0,0.95);
  \draw[grav] (-0.4,0) -- (-0.4,0.55);
  \draw[grav] (0.4,0) -- (0.4,0.55);
  \draw[wline] (-1.0,0) -- (1.0,0);
  \node[mdot,label={[vlab]below:$M$}] at (-0.4,0) {};
  \node[mdot,label={[vlab]below:$M$}] at (0.4,0) {};
\end{tikzpicture}
\;+\cdots\ ,
\end{split}
\label{eq:DefMWorldblob}
\end{equation}
where the first diagram describes scattering by a bulk three-graviton vertex attached to the worldline mass monopole, while the second uses the recoil vertex on the worldline. The third diagram is an example of a one-loop contribution at order $M^2$, with two mass insertions. The ellipsis denotes the remaining diagrams at this order and all higher-order in $M$ corrections. The computation of the point-particle Compton amplitude for nonspinning bodies has recently been pushed to a very high PM order~\cite{Bjerrum-Bohr:2025bqg,Bautista:2026qse,Bjerrum-Bohr:2026fhs,Bautista:2026g4,Brunello:2026c4,Brunello:2026c5,Correia:2026utp}, with several of these results checked against black-hole perturbation theory. These computations become increasingly involved at higher orders, but we essentially skip computing them in this work using the subtraction trick introduced below in Eq.~\eqref{eq:key}, and connecting the remaining part to the tail diagrams from Ref.~\cite{Edison:2023qvg,Edison:2024owb}.

The second part of the transition operator, involving quadrupolar tides, is
\begin{equation}
iT_{\rm tid}=
\begin{tikzpicture}[baseline=0.35cm]
  \draw[grav] (-0.95,0.95) -- (-0.45,0);
  \draw[grav] (0.45,0) -- (0.95,0.95);
  \draw[wline] (-0.85,0) -- (-0.45,0);
  \draw[qprop] (-0.45,0) -- (0.45,0);
  \draw[wline] (0.45,0) -- (0.85,0);
  \node[qdot,label={[vlab]below:$Q$}] at (-0.45,0) {};
  \node[qdot,label={[vlab]below:$Q$}] at (0.45,0) {};
\end{tikzpicture}
\;+\;
\begin{tikzpicture}[baseline=0.35cm]
  \draw[grav] (-1.15,0.95) -- (-0.7,0.6);
  \draw[grav] (-0.7,0.6) -- (-0.3,0);
  \draw[grav] (-0.95,0) -- (-0.7,0.6);
  \draw[grav] (0.45,0) -- (0.95,0.95);
  \draw[wline] (-1.4,0) -- (-0.3,0);
  \draw[qprop] (-0.3,0) -- (0.45,0);
  \draw[wline] (0.45,0) -- (0.85,0);
  \node[blob] at (-0.95,0) {$M$};
  \node[qdot,label={[vlab]below:$Q$}] at (-0.3,0) {};
  \node[qdot,label={[vlab]below:$Q$}] at (0.45,0) {};
\end{tikzpicture}
\;+\cdots\ .
\end{equation}
Here, the first diagram represents the leading tidal contribution, with two graviton-quadrupole vertices connected by a quadrupole propagator. The second diagram is a correction to this, involving the incoming graviton interacting with the mass. The ellipsis denotes the outgoing graviton interacting with the mass as well as all other higher-order in $M$ corrections.

The simple geodesic action above does not yield a finite Compton amplitude at all PM orders. We have to consider ``far-zone'' divergences starting from $(M\omega)^7$~\cite{Saketh:2023bul}, which require tidal counter terms. However, this complication can be addressed simply by adding a minimal set of counter terms and including them in the definition of a ``point particle'', thus rendering $T_{\rm pp}$ finite. These counter terms should only depend on the scale $M$, so that they are universal. In fact, a very natural definition is obtained by choosing the counter terms to be those of a black hole\footnote{A black hole has both conservative and dissipative tidal response, with the latter due to the horizon. Only the conservative tidal-response coefficients are relevant here.}. Any such tidal contributions arising from $T_{\rm pp}$ will appear and scale as $(M\omega)^7(1+\mathcal{O}(M\omega)^2+\dots)$, while that to $T_{\rm tid}$ of a neutron star is known to scale as $(R\omega)^5(1+ \mathcal{O}(M\omega)+\dots)$. Thus, there is a parametric suppression of five powers of compactness $C=M/R$, allowing us to safely neglect the tidal contributions from $T_{\rm pp}$.

Thus, we can define the tidal-scattering phase operator $\Delta_{\rm tid}$ as
\begin{equation}
\begin{aligned}
e^{2i\Delta_{\rm tid}}&\equiv e^{2i\Delta}e^{-2i\Delta_{\rm pp}}=1+iT_{\rm tid}+T_{\rm tid}T_{\rm pp}^\dagger\ .
\end{aligned}
\label{eq:key}
\end{equation}
To constrain the tidal response function using this, we need to compute the eigenvalues of the above operator in the complete description. The eigenvalue of any such $\Delta$ is the scattering phase. We can compute this using a combination of analytic and numerical methods for any known spherically symmetric compact object in general relativity. In particular, we can do this for a neutron star, and then for a black hole, and simply identify
\begin{equation}
\begin{aligned}
\langle\omega'\ell mP|\,\Delta_{\rm tid}\,|\omega\ell mP\rangle&=2\pi\delta(\omega-\omega')\,\delta^{\rm tid}_{\ell\omega},\\
\delta^{\rm tid}_{\ell\omega}&=\delta^{\rm NS}_{\ell\omega}-\Re\,\delta^{\rm BH}_{\ell\omega},
\end{aligned}
\end{equation}
where we take the real part of the black-hole's phase to discard the leading dissipative contribution due to the presence of the event horizon, rendering the tidal scattering phase unitary. The relationship is written in a spherical basis with fixed parity, and we specifically focus on the scattering phase for $\ell=2,~P=+$ modes in this work. This encodes the dominant, electric, quadrupolar tidal response.

The remaining task is then to simply compute the tidal-scattering phase in EFT and perturbation theory and match the results to fix $F(\omega)$. The perturbation theory side is well described by coupled metric-matter perturbation equations in the interior of a neutron star~\cite{Detweiler1985}, and the Regge-Wheeler (RW) equation in the exterior of a neutron star (as well as a black hole). The Regge-Wheeler equation can be efficiently solved by the   Mano-Suzuki-Takasugi (MST) strategy~\cite{Mano:1996gn,Mano:1996mf,Casals:2015nja}. The treatment remains unchanged from Ref.~\cite{Saketh:2026trm}, and we briefly summarize it here.

In the stellar interior, the background metric and configuration is obtained by solving the Tolman–Oppenheimer–Volkoff equations, together with the equation of state. We then include linear, polar, metric and fluid perturbations. In the Regge-Wheeler gauge, the metric perturbations are described by the functions $H_0$, $H_1$, $H_2$ and $K$, and the fluid by the radial and tangential components $W$ and $V$ of the Lagrangian displacement~\cite{Thorne1967,Detweiler1985}. Following Ref.~\cite{Detweiler1985}, the Einstein equations at linear perturbation order reduce to four first-order equations. Given an equation of state and a relation between the Lagrangian variation of pressure and density, these can be solved by demanding regularity at the center, and the metric perturbations can then be matched at the surface to the exterior solution. The adiabatic sound speed relating the Lagrangian variations in pressure and density to each other is computed at fixed composition, as opposed to at equilibrium, since the Urca reaction rates in cold $npe\mu$ matter are slow compared to the anticipated perturbation timescales. The resulting composition stratification provides the buoyancy that supports low-frequency gravity ($g$-) modes.

Outside the star, the polar metric perturbation equations can be reduced to a single, second-order, differential equation---the Regge-Wheeler equation~\cite{PhysRev.108.1063}, satisfied by the RW scalar $\psi_{\rm RW}(t,r,\theta,\varphi) = e^{-i\omega t}Y_{\ell m}(\theta,\varphi)\phi(r)$ with the radial behaviour determined by
\begin{equation}
\frac{\diff^2\phi}{\diff r_*^2}+\left[\omega^2-f\left(\frac{\ell(\ell+1)}{r^2}-\frac{6M}{r^3}\right)\right]\phi=0\ ,
\end{equation}
where $f=1-2M/r$ and $\diff r_*/\diff r=f^{-1}$. At $r=R$, $\phi$ and $\diff\phi/\diff r$ follow from $H_0$ and $K$ through  relations given in Ref.~\cite{Saketh:2026trm}.

The radial equation can be solved as $\phi=B_\nu X_0^\nu+B_{-\nu-1}X_0^{-\nu-1}$, where $X_0^{\nu}$ and $X_0^{-\nu-1}$ are the MST solutions~\cite{Mano:1996mf,Sasaki:2003xr,Casals:2015nja,Saketh:2026trm}, and $\nu$ is a renormalized angular momentum, which reduces to $\ell$ as $\omega\to0$. Matching with the metric perturbations at the stellar surface fixes the ratio $B_{-\nu-1}/B_\nu$, which encodes the internal structure of the star, while $B_{-\nu-1}=B_\nu$ for a black hole

To obtain the scattering phase, we need to take the asymptotic limit $r\gg \omega^{-1}$, but the solutions $X_0^{\nu}$ and $X_0^{-\nu-1}$---which are given as near-zone expansions---do not converge in that regime. They can, however, be related to two different solutions which have convergent expansions in the wave-zone. These are defined as $X_0^\nu = K_\nu X_C^\nu$, and $X_0^{-\nu-1}=K_{-\nu-1}X_C^{-\nu-1}$. Finally, in the wave zone, we can write
\begin{equation}
    \phi = A_{\ell \omega}^{\rm in}e^{-i\omega r_*}+ A_{\ell \omega}^{\rm out}e^{i\omega r_*},
\end{equation}
and identify $e^{2i\delta^{\rm NS}_{\ell\omega}}=(-1)^{\ell+1}A^{\rm out}_{\ell\omega}/A^{\rm in}_{\ell\omega}$ from the amplitudes of the outgoing and incoming waves at infinity.

The tidal scattering phase can now be obtained by subtracting the real part of the phases of a neutron star and black hole, as described before. Common terms cancel in $\delta^{\rm tid}_{\ell\omega}$, and we obtain~\cite{Saketh:2026trm}

\begin{equation}
\begin{split}
2i\delta^{\rm tid}_{\ell\omega}&=\log\mathcal{A}(\mathcal{K}_{\rm NS})-i\,\Im\log\mathcal{A}(\mathcal{K}_{\rm BH})\ ,\\
\mathcal{A}(\mathcal{K})&=\frac{1+ie^{i\pi\nu}\mathcal{K}}{1-ie^{-i\pi\nu}\dfrac{\sin[\pi(\nu+2i\bar\omega)]}{\sin[\pi(\nu-2i\bar\omega)]}\mathcal{K}}\ ,
\end{split}
\label{eq:SPTphase}
\end{equation}
with $\mathcal{K}_{\rm NS}=B_{-\nu-1}K_{-\nu-1}/(B_\nu K_\nu)$ and $\mathcal{K}_{\rm BH}=K_{-\nu-1}/K_\nu$. In practice, we evaluate Eq.~\eqref{eq:SPTphase} numerically for a range of real frequencies, and the resulting $\delta^{\rm tid}_{2\omega}$ provides the input to the matching in Sec.~\ref{sec:ntider}. The same interior solver also gives the quasinormal-mode frequencies of the star, which we use for the oscillator model in Sec.~\ref{sec-osc}. We now turn to the EFT computation.

\section{EFT computation to $M^4$}
\label{sec-resum}
In Eq.~\eqref{eq:key}, the role of the third term is mainly to maintain unitarity of the expression and subtract away infrared divergences that add an overall (divergent) phase which is the same for neutron stars and black holes. Thus, we can focus on the computation of $iT_{\rm tid}$ in a partial-wave basis and then obtain the tidal phase by just dropping infrared divergences and imposing unitarity.

A general diagram contributing to $iT_{\rm tid}$ is contained in the resummation
\begin{equation}
\begin{split}
iT_{\rm tid}&=
\begin{tikzpicture}[baseline=0.1cm]
  \draw[wline] (-0.55,0) -- (0.8,0);
  \draw[wline] (1.45,0) -- (2.8,0);
  \node[blob] (m1) at (0,0) {$M$};
  \node[blob] (m2) at (2.25,0) {$M$};
  \draw[grav] (-0.75,0.95) -- (m1.north west);
  \draw[grav] (m1.north east) to[out=60,in=110] (0.8,0);
  \draw[qprop] (0.8,0) -- (1.45,0);
  \draw[grav] (1.45,0) to[out=70,in=120] (m2.north west);
  \draw[grav] (m2.north east) -- (3.0,0.95);
  \node[qdot,label={[vlab]below:$Q$}] at (0.8,0) {};
  \node[qdot,label={[vlab]below:$Q$}] at (1.45,0) {};
\end{tikzpicture}\\
&+
\begin{tikzpicture}[baseline=0.1cm]
  \draw[wline] (-0.55,0) -- (0.8,0);
  \draw[wline] (3.2,0) -- (4.55,0);
  \node[blob] (m1) at (0,0) {$M$};
  \node[blob] (m3) at (4.0,0) {$M$};
  \node[blob] (m2) at (2.0,0.5) {$M$};
  \draw[grav] (-0.75,0.95) -- (m1.north west);
  \draw[grav] (m1.north east) to[out=60,in=110] (0.8,0);
  \draw[qprop] (0.8,0) -- (1.45,0);
  \draw[grav] (1.45,0) to[out=60,in=200] (m2.west);
  \draw[grav] (m2.east) to[out=-20,in=120] (2.55,0);
  \draw[qprop] (2.55,0) -- (3.2,0);
  \draw[grav] (3.2,0) to[out=70,in=120] (m3.north west);
  \draw[grav] (m3.north east) -- (4.75,0.95);
  \node[qdot,label={[vlab]below:$Q$}] at (0.8,0) {};
  \node[qdot,label={[vlab]below:$Q$}] at (1.45,0) {};
  \node[qdot,label={[vlab]below:$Q$}] at (2.55,0) {};
  \node[qdot,label={[vlab]below:$Q$}] at (3.2,0) {};
\end{tikzpicture}
+\cdots\ .
\end{split}
\end{equation}
This resummation has two parts: the first is the vertex correction due to propagation of waves in the curved background before or after interacting with the tidal quadrupole. This is denoted by the $M$-blob on the worldline defined in Eq.~\eqref{eq:DefMWorldblob}.
The second part is the dressing of the tidal-quadrupole propagator due to the mass. The second diagram in the above equation is the first such correction with the $M$-blob in between the quadrupole propagators defined as
\begin{equation}
\begin{split}
\begin{tikzpicture}[baseline=-0.1cm]
  \node[blob] (m) at (0.6,0.5) {$M$};
  \draw[grav] (0,0) to[out=60,in=200] (m.west);
  \draw[grav] (m.east) to[out=-20,in=120] (1.2,0);
  \node[qdot,label={[vlab]below:$Q$}] at (0,0) {};
  \node[qdot,label={[vlab]below:$Q$}] at (1.2,0) {};
\end{tikzpicture}
=
\begin{tikzpicture}[baseline=-0.1cm]
  \draw[grav] (0,0) to[out=75,in=105] (0.9,0);
  \node[qdot,label={[vlab]below:$Q$}] at (0,0) {};
  \node[qdot,label={[vlab]below:$Q$}] at (0.9,0) {};
\end{tikzpicture}
+
\begin{tikzpicture}[baseline=-0.1cm]
  \draw[grav] (0,0) to[out=75,in=180] (0.55,0.5);
  \draw[grav] (0.55,0.5) to[out=0,in=105] (1.1,0);
  \draw[grav] (0.55,0) -- (0.55,0.5);
  \draw[wline] (-0.15,0) -- (1.25,0);
  \node[qdot,label={[vlab]below:$Q$}] at (0,0) {};
  \node[mdot,label={[vlab]below:$M$}] at (0.55,0) {};
  \node[qdot,label={[vlab]below:$Q$}] at (1.1,0) {};
\end{tikzpicture}
+\cdots \ .
\end{split}
\label{eq:DefQMQblob}
\end{equation}
We note that the quadrupole propagator dressing also encodes the absolute value of the vertex corrections through the optical theorem, including ultraviolet (UV) divergences.

At order $M^2$, we find that the dressed propagator develops UV divergences which cannot be regulated by a real-valued tidal response $F(\omega)$ (see Eq.~\eqref{eq:Qprop}). This can be seen from the fact that the vertex corrections are also UV divergent starting from order $M^2$, and thus a finite dressed propagator will produce a UV divergent $i T_{\rm tid}$. To obtain a finite amplitude, we need to also renormalise the quadrupole-tidal-field coupling, and replace the interaction term in the action in such a way that
\begin{equation}
\begin{gathered}
-\frac12\int \diff \tau\,Q^{\mu\nu}E_{\mu\nu}\;\to\;-\frac12\int\frac{\diff \omega}{2\pi}\,g_{\rm B}(\omega)\,\tilde Q^{\mu\nu}(\omega)\,\tilde E_{\mu\nu}(-\omega),
\end{gathered}
\end{equation}
where
\begin{equation}
    g_{\rm B}(\omega)=1-\bar\omega^2 \kappa_{\rm B}+\bar{\omega}^4\sigma_{\rm B}+\mathcal O(\bar\omega^6)
    \label{eq:DefgB}
\end{equation}
is a bare coupling constant that allows for a series of local derivative couplings along the worldline. Note that we do not include terms odd in the expansion parameter $\bar\omega$ since such terms are odd under time-reversal, and must not appear in the time-reversal invariant, conservative sector considered here.

With this modification, we can generically write the dressed propagator as:
\begin{equation}
\begin{split}
&\big\langle \mathbb{T}\tilde{Q}^{\mu\nu}(\omega)\tilde{Q}^{\rho\sigma}(\omega')\big\rangle\big|_{\rm dressed}\\
&=
\begin{tikzpicture}[baseline=-0.1cm]
  \draw[qprop] (0,0) -- (0.65,0);
  \node[qdot,label={[vlab]below:$Q$}] at (0,0) {};
  \node[qdot,label={[vlab]below:$Q$}] at (0.65,0) {};
\end{tikzpicture}
+
\begin{tikzpicture}[baseline=-0.1cm]
  \node[blob] (m) at (1.15,0.5) {$M$};
  \draw[qprop] (0,0) -- (0.55,0);
  \draw[grav] (0.55,0) to[out=60,in=200] (m.west);
  \draw[grav] (m.east) to[out=-20,in=120] (1.75,0);
  \draw[qprop] (1.75,0) -- (2.3,0);
  \foreach \x in {0,0.55,1.75,2.3} \node[qdot,label={[vlab]below:$Q$}] at (\x,0) {};
\end{tikzpicture}
+\cdots\\
&=-\frac{iR^5\,\Pi_d^{\mu\nu,\rho\sigma}}{F_{\rm B}^{-1}(\omega)-\frac{(R\omega)^5}{160\pi m_{\rm pl}^2}\,g_{\rm B}(\omega)g_{\rm B}(\omega')\,\Sigma(\omega)}2\pi \delta (\omega +\omega')\ ,
\end{split}
\end{equation}
where we have used
\begin{equation}
\begin{split}
\begin{tikzpicture}[baseline=-0.1cm]
  \node[blob] (m) at (0.6,0.5) {$M$};
  \draw[grav] (0,0) to[out=60,in=200] (m.west);
  \draw[grav] (m.east) to[out=-20,in=120] (1.2,0);
  \node[qdot,label={[vlab]below:$Q$}] at (0,0) {};
  \node[qdot,label={[vlab]below:$Q$}] at (1.2,0) {};
\end{tikzpicture}
&=\frac{i\omega^5}{160\pi m_{\rm pl}^2}\,g_{\rm B}(\omega)g_{\rm B}(\omega')\,\Sigma(\omega)\,\Pi_d^{\mu\nu,\rho\sigma},\\
\Sigma&=\Sigma_{(0)}+\bar\omega\,\Sigma_{(1)}+\bar\omega^2\,\Sigma_{(2)}+\cdots.
\end{split}
\label{eq:dressing}
\end{equation}
Note that the subscript $\rm B$ in all the objects above denotes that they are bare. The problem now boils down to simply computing the dimensionless quadrupole self-energy $\Sigma$.

\subsection{Self-energy correction up to 4PM}
Here we use the results computed in the context of radiation reaction of binaries in various works~\cite{Edison:2024owb,Edison:2023qvg,Galley2016} to compute the self energy corrections up to 4PM. The relevance of such computations to the problem of compact-object tides has, to our knowledge, not been previously recognized.

At order $M^0$, we have
\begin{equation}
\begin{split}
    \Sigma_{(0)} = i\bigg[&1-\epsilon\left(L-\frac{9}{10}\right)\\
    &+\epsilon^2\left(\frac{L^2}{2}-\frac{9}{10}L+\frac{43}{50}+\frac{\pi^2}{4}\right)\bigg]+\mathcal O(\epsilon^3) \ ,
\end{split}
\end{equation}
where
\begin{equation}
    L\equiv\ln\frac{4e^{2\gamma_{\rm E}}\omega^2}{\mu^2}-i\pi
    \label{eq:DefL}
\end{equation}
This arises from the first term in Eq.~\eqref{eq:DefQMQblob}, which is a one-loop diagram that corresponds to radiation reaction \cite{Galley:2009px,Thorne:1969rba,Burke:1970wx}. Beyond this, starting from order $M^1$, we have two or more loops. However, the imaginary part of $\Sigma_{(1)}$ can be obtained simply from the optical theorem and unitarity as shown in Ref.~\cite{Saketh:2026trm}, and we have
\begin{equation}
    \Im \Sigma_{(1)} = 2\pi .
\end{equation}
The tidal-scattering phase arising from these two contributions was obtained previously in Ref.~\cite{Saketh:2026trm}.

The self-energy at order $M^1$ is given by
\begin{equation}
\begin{split}
    \Sigma_{(1)}&=\frac{1}{\epsilon}-2L+\frac{41}{15}\\
    &+\epsilon\Big(2L^2-\frac{82}{15}L+\frac{\pi^2}{2}+\frac{3667}{450}\Big)+\mathcal O(\epsilon^2)\ \ ,
\end{split}
\label{eq:fullM}
\end{equation}
where $L$ is defined in Eq.~\eqref{eq:DefL}.
As we can see, the self-energy at 1PM order is ultraviolet divergent. This divergence can be regulated by absorbing it into the bare tidal response using
\begin{equation}
    F_{\rm B}^{-1}(\omega)=F^{-1}(\omega;\mu)+\frac{\bar\omega\,(R\omega)^5}{160\pi m_{\rm pl}^2\,\epsilon} \ ,
    \label{eq:linM}
\end{equation}
where $F(\omega;\mu)$ is the finite, renormalised tidal response, whose beta function, i.e., the renormalization-group flow equation, can be obtained by a derivative with respect to the scale $\mu$:
\begin{equation}
    \mu\frac{\diff F^{-1}}{\diff\mu}=\frac45\,\bar\omega\,(R\omega)^5\ .
    \label{eq:BetaLinOmega}
\end{equation}
The above originates from the tail effect \cite{Blanchet:1992br,Blanchet:1997ji}, in which the field radiated by the quadrupole scatters off the mass $M$ before it returns to the worldline.

At order $M^2$, the self-energy is given by
\begin{equation}
\begin{aligned}
    \Sigma_{(2)}=&i\frac{107}{105}\left[\frac1\epsilon-3L\right]+i\Big(\frac{675359}{44100}+\frac{4\pi^2}{3}\Big)\\
    &+i\epsilon\bigg[\frac{321}{70}L^2-\Big(\frac{675359}{14700}+4\pi^2\Big)L\\
    &+\frac{1259125247}{9261000}+\frac{3449\pi^2}{420}-40\zeta(3)\bigg]+\mathcal O(\epsilon^2).
\end{aligned}
\end{equation}
Once again, we find a divergence, but the divergence is now imaginary. As mentioned previously, this is actually associated with divergent vertex corrections, and can be regulated by using the quadratic coefficient of $g_{\rm B}(\omega)$ as
\begin{equation}
    g_{\rm B}(\omega)=1-\bar\omega^2\left[\kappa(\mu)+\frac{107}{210\,\epsilon}\right]+\mathcal O(\bar\omega^3)\ ,
    \label{eq:gsub}
\end{equation}
where $\kappa(\mu)$ is the finite quadratic coupling whose bare version $\kappa_{\rm B}$ is defined in Eq.~\eqref{eq:DefgB}. This gives the beta function
\begin{equation}
    \mu\frac{\diff \kappa}{\diff\mu}=\frac{214}{105} \ .
    \label{eq:BetaFunckappa}
\end{equation}
This beta function is universal and the same coefficient governs the logarithmic running of the radiative mass quadrupole in the PN expansion~\cite{Goldberger:2009qd,Mandal:2023hqa}.
The value of $\kappa$ at a reference scale, by contrast, depends on the star. The EFT does not determine it, so it must be fixed by matching, and we obtain $\kappa(R^{-1})$ from the stellar scattering phase in Sec.~\ref{sec-osc}.

Next,  we quote below the result for the 3PM contribution to the bare self-energy
\begin{align}
\Sigma_{(3)}
=&\frac{214}{105}\Bigg[
\frac{1}{4\epsilon^2}
-\frac{1}{\epsilon}
 \left(L-\frac{252583}{59920}\right)
+2L^2-\frac{252583}{14980}L
\nonumber\\
&-\frac{5\pi^2}{12}
-\frac{840}{107}\zeta(3)
+\frac{1583459537}{37749600}
\Bigg]+\mathcal{O}(\epsilon),
\end{align}
Now, the total result at 3PM, i.e., using $g_{\rm B}(\omega)g_{\rm B}(-\omega)\Sigma(\omega)$ yields
\begin{equation}
    -\frac{107}{210\epsilon^2}+\frac{\frac{512077}{88200}-2 \kappa(\mu)}{\epsilon }+\Sigma_{\rm R}^{\rm 3PM},
\end{equation}
where
\begin{equation}
\begin{aligned}
\Sigma_{\rm R}^{\rm 3PM}
&
=\bigg[
\frac{214}{105}L^{2}
-\frac{634913}{22050}L
+4\kappa(\mu)\left(L-\frac{41}{30}\right)\\
&\qquad+\frac{1429650889}{18522000}
-\frac{428\pi^{2}}{315}
-16\zeta(3)
\bigg]\ ,
\end{aligned}
\label{eq:SigmaR3PM}
\end{equation}
Unlike the imaginary 2PM pole, these poles are real, like the 1PM pole in Eq.~\eqref{eq:fullM}. They cannot be removed through the coupling. Time-reversal invariance forbids a $\bar\omega^3$ term in $g_{\rm B}$. Moreover, any new real coefficient in $g_{\rm B}$ enters first through its product with $\Sigma_{(0)}=i$, so it can only cancel imaginary poles. We therefore absorb the real 3PM poles into the tidal response $F$ by rewriting Eq.~\eqref{eq:linM} as
\begin{equation}
\begin{split}
        F_{\rm B}^{-1}(\omega) &=F^{-1}(\omega;\mu)+\frac{\bar\omega\,(R\omega)^5}{160\pi m_{\rm pl}^2\,\epsilon} \\&- \frac{\bar{\omega}^3(R\omega)^5}{160 \pi m_{\rm pl}^2}\left(\frac{107}{210 \epsilon^2}-\frac{\frac{512077}{88200}-2 \kappa(\mu)}{\epsilon }\right),
\end{split}
\end{equation}
which modifies Eq.~\eqref{eq:BetaLinOmega} to
\begin{equation}
    \mu\frac{\diff F^{-1}}{\diff\mu}=\frac45\,\bar\omega\,(R\omega)^5-\frac85\Big(\kappa(\mu)-\frac{512077}{88200}\Big)\bar\omega^3(R\omega)^5\ ,
    \label{eq:BetaLinCubeOmega}
\end{equation}
At this order, the running of $F^{-1}$ depends on $\kappa(\mu)$, which itself runs according to Eq.~\eqref{eq:BetaFunckappa}, so the two flows must be solved together.

The bare self-energy at 4PM order takes the form
\begin{align}
\Sigma_{(4)}
&=i\frac{22898}{11025}
\Bigg[
\frac{1}{4\epsilon^2}
-\frac{1}{2\epsilon}
\left(
\frac{5}{2}L
-\frac{840}{107}\zeta(2)
-\frac{5249287}{343470}
\right)
\nonumber\\
&\qquad\qquad+\frac{25}{8}L^2
-\frac{5}{2}
\left(
\frac{840}{107}\zeta(2)
+\frac{5249287}{343470}
\right)L\nonumber\\
&\qquad\qquad+\frac{7259215}{91592}\zeta(2)-\frac{2940}{107}\zeta(3)
-\frac{176400}{11449}\zeta(4)\nonumber\\
&\qquad\qquad+\frac{90973869743}{673201200}
\Bigg]+\mathcal{O}(\epsilon)\ .
\end{align}
Although $\Sigma_{(4)}$ contains an overall factor of $i$, it is not purely imaginary since $L$ contains an imaginary piece. After including the lower-order counterterms, we find that all the poles in $\Sigma_{(4)}$, including their real parts due to the imaginary piece in $L$, cancel. The divergences that remain at this order are absorbed by extending Eq.~\eqref{eq:gsub} to
\begin{equation}
\begin{split}
    g_{\rm B}(\omega)&=1-\bar\omega^2\Big(\kappa(\mu)+\frac{107}{210\,\epsilon}\Big)+\bar\omega^4\bigg[\sigma(\mu)\\
    &+\frac{107\,\kappa(\mu)}{210\,\epsilon}+\frac{11449}{88200\,\epsilon^2}-\frac{1695233}{4630500\,\epsilon}\bigg]+\mathcal O(\bar\omega^6)\ ,
\end{split}
\label{eq:gsub4}
\end{equation}
where $\sigma(\mu)$ is the finite quartic coupling whose bare version is defined in Eq.~\eqref{eq:DefgB}. Its beta function can be read off by differentiating the above equation and using Eq.~\eqref{eq:BetaFunckappa} for the beta function of the quadratic-in-$\bar\omega$ coupling $\kappa$. This gives
\begin{equation}
    \mu\frac{\diff\sigma}{\diff\mu}=\frac{214}{105}\,\kappa(\mu)-\frac{3390466}{1157625}\ .
    \label{eq:sigmaflow}
\end{equation}
Substituting Eq.~\eqref{eq:gsub4} into $g_{\rm B}(\omega)g_{\rm B}(-\omega)\Sigma(\omega)$ cancels all the 4PM poles, and we obtain the 4PM contribution to the renormalised self energy to be
\begin{align}
\Sigma_{\rm R}^{\rm 4PM}
={}&i\Bigg[
\frac{22898}{11025}L^2
-\left(
\frac{8301847}{257250}
+\frac{1712}{105}\zeta(2)
\right)L
\nonumber\\
&\quad+\frac{275977249}{1944810}
+\frac{1269826}{11025}\zeta(2)
-\frac{1712}{105}\zeta(3)
\nonumber\\
&\quad-32\zeta(4)+2\kappa(\mu)
\left(
\frac{214}{105}L
-\frac{634913}{44100}
-\frac{4\pi^2}{3}
\right)\nonumber\\
&\quad+\kappa(\mu)^2+2\sigma(\mu)
\Bigg].
\label{eq:SigmaR4PM}
\end{align}

\begin{figure*}[t!]
    \centering
    \begin{subfigure}[t]{0.5\textwidth}
        \centering
        \includegraphics[width=\textwidth]{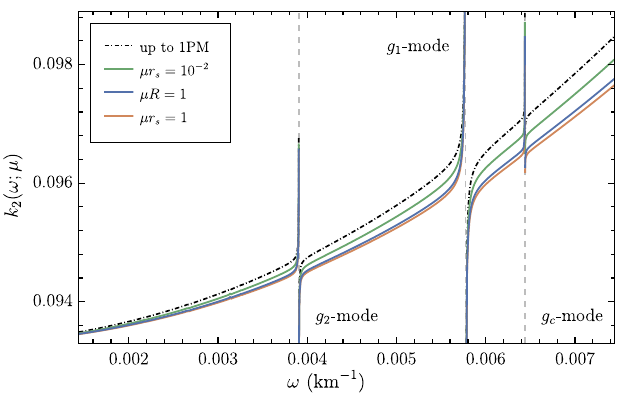}
        \caption{At low frequencies, across the $g_2$, $g_1$ and $g_c$ resonances}
    \end{subfigure}~
    \begin{subfigure}[t]{0.5\textwidth}
        \centering
        \includegraphics[width=\textwidth]{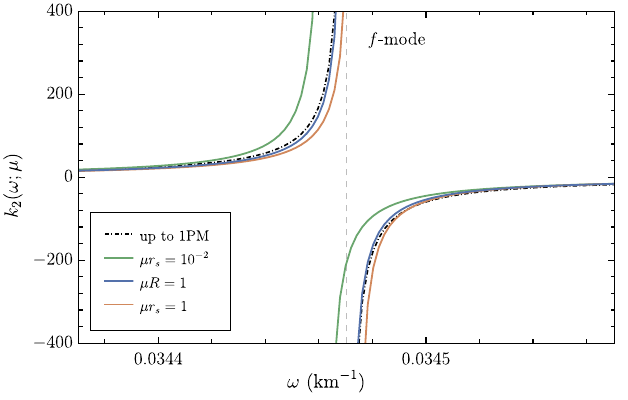}
        \caption{Close to the $f$-mode resonance}
        \label{fig:near_f}
    \end{subfigure}
    \caption{The renormalized tidal response $k_2(\omega;\mu)$ for a $1.4\,M_\odot$ BSk22 star near the $g_2$, $g_1$ and $g_c$ modes (a) and near the $f$-mode (b), for $\mu=R^{-1}$, $\mu r_s=1$ and $\mu r_s=0.01$. The dash-dotted black curve, labelled ``up to 1PM'', is the response of Ref.~\cite{Saketh:2026trm}, which does not include the higher-PM corrections computed here, and the dashed vertical lines mark the mode frequencies. The curves follow directly from the numerically computed tidal phase through Eq.~\eqref{eq:tidmain}, with interpolation on the values on the numerical frequency grid.}
    \label{fig:near_g}
\end{figure*}

We can thus finally write the gravitationally-dressed, 4PM-propagator as
\begin{equation}
\begin{split}
&\big\langle \mathbb{T}\tilde{Q}^{\mu\nu}(\omega)\tilde{Q}^{\rho\sigma}(\omega')\big\rangle\big|_{\rm dressed}\\
&=
-\frac{iR^5\,\Pi_d^{\mu\nu,\rho\sigma}}{F^{-1}(\omega;\mu)-\frac{(R\omega)^5}{160\pi m_{\rm pl}^2}\,\Sigma_{\rm R}(\omega;\mu)}2\pi \delta (\omega +\omega')\ ,
\end{split}
\label{eq:QQdressed}
\end{equation}
where $\Sigma_{\rm R}$ is the renormalized self-energy. It is the finite remainder of $g_{\rm B}(\omega)g_{\rm B}(-\omega)\Sigma(\omega)$ after the counterterms in $F_{\rm B}$ and $g_{\rm B}$ have cancelled all poles and $\epsilon\to0$. Collecting the finite parts found above order by order, we have
\begin{equation}
    \begin{split}
        \Sigma_{\rm R}&=i-2\bar\omega\left(L-\frac{41}{30}\right)\\
        &-i \bar\omega^2\bigg[2\kappa(\mu)+\frac{214}{105}L-\frac{634913}{44100}-\frac{4\pi^2}{3}\bigg] \\& + \bar\omega^3\Sigma^{\rm 3PM}_{\rm R}+\bar\omega^4\Sigma^{\rm 4PM}_{\rm R}\ .
    \end{split}
\end{equation}
The first two terms reproduce the result of Ref.~\cite{Saketh:2026trm}. The constant $i$ is the leading-order radiation reaction on the quadrupole, and the imaginary part of the 1PM term, $2\pi\bar\omega$ from the $-i\pi$ in $L$, is the tail correction obtained above from the optical theorem. The 2PM, 3PM and 4PM terms are new. Their explicit logarithms $L$ carry a dependence on $\mu$ that is compensated by the running of $F$, $\kappa$ and $\sigma$ in Eqs.~\eqref{eq:BetaLinCubeOmega}, \eqref{eq:BetaFunckappa} and \eqref{eq:sigmaflow}. In the next subsection, we use this propagator to extract $F(\omega;\mu)$ from the scattering phase.

\subsection{Tidal response to 4PM}
\label{sec:ntider}

We now relate the renormalized response to the tidal scattering phase. After dropping the infrared-divergent phase since it cancels out in the subtraction~\cite{Ivanov:2026icp} and imposing unitarity, the phase follows directly from the dressed propagator in Eq.~\eqref{eq:QQdressed}.
Thus, the tidal scattering phase to order $M^4$ is given by
\begin{equation}
\begin{gathered}
    e^{2i\delta^{\rm tid}_{\ell\omega}}=\frac{F^{-1}(\omega;\mu)-\frac{(R\omega)^5}{5}\Sigma_{\rm R}^*(\omega;\mu)}{F^{-1}(\omega;\mu)-\frac{(R\omega)^5}{5}\Sigma_{\rm R}(\omega;\mu)}\ ,\qquad \ell=2\ ,\\
\end{gathered}
\label{eq:sfin}
\end{equation}
The presence of several undetermined quantities on the right-hand side, the function $F(\omega;\mu)$ and the couplings $\kappa(\mu)$ and $\sigma(\mu)$ inside $\Sigma_{\rm R}$, may make it unclear how to proceed. However, we can resolve this ambiguity by redefining the quadrupole moment in the action to absorb finite-vertex corrections into $F(\omega;\mu)$, leaving behind terms relevant only at higher PM/PN orders. Equivalently, we can redefine our tidal response by the substitution
\begin{equation}
        F(\omega;
    \mu)\rightarrow \frac{F(\omega;\mu)}{\big[1-\bar{\omega}^2\kappa(\mu)+\bar{\omega}^4\sigma(\mu)\big]^2}\ ,
    \label{eq:fsub}
\end{equation}
which does not modify Eq.~\eqref{eq:sfin} to $\mathcal{O}(\bar{\omega}^5)$ except removing $\kappa(\mu)$ and $\sigma(\mu)$.

Therefore, we can simply set $\kappa(\mu)=\sigma(\mu) = 0$, and invert the first expression in Eq.~\eqref{eq:sfin} to obtain the tidal response. However,  this step modifies the beta function and we obtain
\begin{equation}
\begin{split}
        \mu \frac{\diff F}{\diff\mu} &= -F^2\bar{\omega}\,(R\omega)^5\left(\frac{4}{5}+\frac{512077}{55125}\bar{\omega}^2\right)\\
        &\quad-\left(\frac{428}{105}+\frac{6780932}{1157625}\,\bar{\omega}^2\right)\bar{\omega}^2 F\ .
\end{split}
\end{equation}
Note that this is analogous to the running equation quoted in  recent Born-series and Sommerfeld-resummation papers~\cite{Caron-Huot:2025tlq,Chang:2026eti,Correia:2026utp}, except that it is missing the constant term on the right-hand side. The constant term arises from far-zone divergences which we include in the point-particle sector here.

Finally, we switch normalization to write the expression for $k_2(\omega;\mu) = (3/4)F(\omega;\mu)$, which is generally used in literature\footnote{With this normalization, $k_2$ reduces at $\omega\to0$ to the usual static Love number, $k_2^{\rm s}=3\lambda/(2R^5)$, where $\lambda$ is the static tidal deformability. For the oscillator model, $\lambda=\sum_n\lambda_n$, see Eqs.~\eqref{eq:Fosc} and~\eqref{eq:Seffad}.}, as
\begin{equation}    k_2(\omega;\mu)=\frac{15\tan\delta^{\rm tid}_{2\omega}}{4(R\omega)^5\big[\Im\Sigma_{\rm R}(\omega;\mu)+\Re\Sigma_{\rm R}(\omega;\mu)\tan\delta^{\rm tid}_{2\omega}\big]}\ ,
    \label{eq:tidmain}
\end{equation}
with the flow equation
\begin{equation}
\begin{split}
        \mu \frac{\diff k_2}{\diff \mu} &= -k_2^2\bar{\omega}\,(R\omega)^5\left(\frac{16}{15}+\frac{2048308}{165375}\bar{\omega}^2\right)\\
        &\quad-\left(\frac{428}{105}+\frac{6780932}{1157625}\,\bar{\omega}^2\right)\bar{\omega}^2k_2\ .
\end{split}
    \label{eq:k2flow}
\end{equation}
The only thing left to choose is a renormalization scale, and a simple choice is $\mu=R^{-1}$, which may be regarded loosely as a proxy for the UV scale of the system.

It is worth noting that the $\bar{\omega}^2$ contribution to the response dominates over the linear-in-$\bar{\omega}$ term when we are away from resonances owing to the large power of $R\omega$, and the smallness of $k_2$. Nevertheless, close to resonances the linear-in-$\bar{\omega}$ term becomes important and is in fact related to the running of resonant mode frequencies and shifts them as a function of $\mu$.

We  can now use Eq.~\eqref{eq:tidmain} to compute $k_2(\omega;\mu)$ as a function of frequency for various values of $\mu$ for the BSk22 model with $M=1.4M_\odot$. The dependence on $\mu$ helps quantify the impact of logarithmic terms on the tidal response. Following Ref.~\cite{Apostolidis:2026qsg}, we vary between $\mu r_s=0.01$ and $\mu r_s=1$, where $r_s=2M$, which includes our choice $\mu=R^{-1}$ as it lies in between ($\mu r_s = r_s/R \approx 0.32$).
We also overlay the tidal response calculated in Ref.~\cite{Saketh:2026trm} to quantify the effect of higher PM order corrections.

The results are shown in Fig.~\ref{fig:near_g}. As in Ref.~\cite{Saketh:2026trm}, the dynamical tidal-response captures $g$-modes and the dominant $f$-mode resonance. Additionally, it appears that the tidal response function generically decreases in magnitude with increasing $\mu$. This follows from the flow equation in Eq.~\eqref{eq:k2flow} which shows a decrease in $k_2$ with $\mu$. For  low frequencies, away from resonances, it is primarily the second term on the right-hand side that determines the result.  However, owing to the $\mu$ dependence, $k_2$ is not an observable by itself. Nevertheless, given that many works incorporate resonances in $k_2(\omega)$ while neglecting curvature corrections that lead to renormalization~\cite{Steinhoff:2016rfi,Andersson:2025iyd}, it is worth taking a look at the renormalized tidal response function and its dependence on $\mu$.

Closer to the $f$-mode, the first term in the flow Eq.~\eqref{eq:k2flow} also becomes important, and since it is a constant flow when written in terms of $k_2^{-1}$, it shifts the resonant frequencies. Fig.~\ref{fig:near_f} shows this shift as $\mu$
is varied. The shifting in the pole locations is not paradoxical since poles of the tidal response function do not directly map to the quasinormal-mode frequencies. In fact, Eq.~\eqref{eq:sfin} tells us that the quasinormal-mode frequencies are instead given by the condition
\begin{equation}
    F^{-1}(\omega;\mu) - \frac{1}{5}(R\omega)^5 \Sigma_{\rm R}(\omega;\mu)= 0\ .
\end{equation}
This follows since quasinormal modes are poles of the scattering matrix with purely outgoing boundary conditions, and by Eq.~\eqref{eq:sfin}, this corresponds to poles of the dressed propagator in Eq.~\eqref{eq:QQdressed}, and not of $F$ alone. Furthermore, note that the above condition is independent of the scale $\mu$. At linear order in $\bar\omega$, this can be gleaned by observing that the term $-2\bar\omega\ln(\omega^2/\mu^2)$ in $\Sigma_{\rm R}$ cancels a similar running logarithm term coming from $F^{-1}$ in Eq.~\eqref{eq:BetaLinOmega}, and likewise at higher orders in $\bar\omega$ through Eq.~\eqref{eq:BetaLinCubeOmega}. Thus, we see that the poles of the tidal response function move with $\mu$, while the quasinormal modes stay fixed.

\section{An action for the quadrupole moment}
\label{sec-osc}

So far, we have shown how a generic tidal response function can be obtained from the scattering phase and how it is renormalized. However, the resulting function is obtained numerically, which is often inconvenient for waveform models.

It is thus worth studying whether a simple approximate action can reproduce the dynamics. To that end, we consider the often studied oscillator model~\cite{Steinhoff:2016rfi,Chakrabarti:2013lua,Gao:2026inprep,Mandal:2023lgy,Steinhoff:2021dsn}, and write the worldline action including an explicit quadrupole action as~\cite{Mandal:2023hqa}
\begin{equation}
\begin{split}
        \mathcal{S} &= \mathcal{S}_{\rm EH}+\mathcal{S}_{\rm GF}- M \int \diff \tau   \\& \quad  +\int \diff \tau \sum_n\frac{1}{4 \lambda_n\omega_n^2}\left[\frac{\Diff Q^{(n)}_{\mu\nu}}{\diff \tau}\frac{\Diff Q_{(n)}^{\mu\nu}}{\diff \tau}-\omega_n^2 Q^{(n)}_{\mu\nu}Q_{(n)}^{\mu\nu}\right]\\& \quad - \int \frac{\diff \omega}{2\pi}\frac{g_{\rm B}(\omega)}{2}\tilde{Q}^{\mu\nu}(\omega)\tilde{E}_{\mu\nu}(-\omega),
\end{split}
\end{equation}
where
\begin{equation}
    Q^{\mu\nu}=\sum_{n}Q^{\mu\nu}_{(n)},~ n=f,~g_1,~g_2,~g_c,\ldots\ .
\end{equation}
We impose $Q^{\mu\nu}u_\nu=0$, which removes the components of $Q^{\mu\nu}$ along the worldline, so that only its spatial components in the rest frame of the star remain, as in Refs.~\cite{Steinhoff:2016rfi}.
 Here $\Diff/\diff\tau$ is the covariant derivative along the worldline.  We therefore replace covariant derivatives by ordinary ones in what follows\footnote{The couplings to the gravitational field it generates are quadratic in $Q^{\mu\nu}$, and can enter the two-point function of $Q^{\mu\nu}$ for an isolated body only through the body's own field on its worldline, which is scaleless and vanishes in dimensional regularization.}. From the above action, it is straightforward to compute the two-point function of the quadrupole:
\begin{equation}
  \big\langle \mathbb{T}\tilde Q^{\mu\nu}(\omega)\tilde Q^{\rho\sigma}(\omega')\big\rangle= -\sum_n\frac{i \Pi_d^{\mu\nu,\rho\sigma}\lambda_n 4\pi \delta(\omega+\omega')}{1-\frac{\omega^2}{\omega_n^2}} \ .
\end{equation}
Matching this expression to Eq.~\eqref{eq:Qprop}, we obtain the tidal-response function of the oscillator, $F_{\rm osc}$;
\begin{equation}
    R^5 F_{\rm osc}(\omega) = \sum_n\frac{2\lambda_n}{1-\frac{\omega^2}{\omega_n^2}}.
    \label{eq:Fosc}
\end{equation}

However, this bare action is insufficient because gravitational dressing leads to self-energy divergences as in Eq.~\eqref{eq:linM}, which requires adding $\mathcal{O}(\omega^6)$ and $\mathcal{O}(\omega^8)$ terms to the response. Additionally, the bare action couples oscillators associated with modes diagonally, but gravitational interactions can couple different oscillators together. These effects can be incorporated by supplementing the oscillator action with the counter-term contribution
\begin{align}
        \mathcal{S}_{\rm CT} &= \int \diff\tau \left[\frac{M}{2}\alpha_{\rm B}\, \dddot{Q}^{\mu\nu} \dddot{Q}_{\mu\nu}+\frac{M^3}{2}\alpha^{(3)}_{\rm B}\, \ddddot{Q}^{\mu\nu} \ddddot{Q}_{\mu\nu}\right]
\end{align}
where the bare couplings are given as,
\begin{align}
    \alpha_B&=\alpha(\mu) - \frac{\tilde{\mu}^{-4\epsilon}}{5\epsilon}\\
    \alpha^{(3)}_B&=\alpha^{(3)}(\mu)+\frac{\tilde{\mu}^{-8\epsilon}}{5}\left(\frac{107}{210\epsilon^2}+\frac{2\kappa(\mu)-\frac{512077}{88200}}{\epsilon}\right)
\end{align}
Their poles are fixed by requiring that they reproduce the counterterms in $F_{\rm B}^{-1}$, Eq.~\eqref{eq:linM} at 1PM and its extension to 3PM in Sec.~\ref{sec-resum}. The finite parts $\alpha(\mu)$ and $\alpha^{(3)}(\mu)$ are new parameters of the model. This choice modifies the bare $ \big\langle \mathbb{T}\tilde Q^{\mu\nu}(\omega)\tilde Q^{\rho\sigma}(\omega')\big\rangle$ response to
\begin{equation}
     \frac{-i R^5\Pi_d^{\mu\nu,\rho\sigma} 2\pi \delta(\omega+\omega')}{F_{\rm osc}(\omega;\mu)^{-1} +\frac{\bar{\omega}(R\omega)^5}{160\pi m_{\rm pl}^2}\Big[\frac{1}{\epsilon}-\bar\omega^2\Big(\frac{107}{210\epsilon^2}+\frac{2\kappa(\mu)-\frac{512077}{88200}}{\epsilon}\Big)\Big]}\ ,
     \label{eq:bareosc3PM}
\end{equation}
where
\begin{equation}
\begin{split}
    F_{\rm osc}(\omega;\mu)^{-1} &=  F_{\rm osc}(\omega)^{-1}- \big[\alpha(\mu)+\alpha^{(3)}(\mu)\,\bar{\omega}^2\big]\bar{\omega}(R\omega)^5\ .
\end{split}
    \label{eq:1PMoscmap}
\end{equation}
The coefficients $\alpha$ and $\alpha^{(3)}$ run according to
\begin{equation}
    \mu\frac{\diff\alpha}{\diff\mu}=-\frac{4}{5},\quad \mu\frac{\diff\alpha^{(3)}}{\diff\mu}=\frac{8}{5}\Big(\kappa(\mu)-\frac{512077}{88200}\Big)\ .
    \label{eq:1PM3PMrun}
\end{equation}

While Eq.~\eqref{eq:bareosc3PM} is divergent, it is divergent in precisely the manner required to cancel the divergences due to gravitational dressing at 1PM and 3PM orders, as evident from Eqs.~\eqref{eq:BetaLinOmega} and \eqref{eq:BetaLinCubeOmega}. Given the quadratic nature of the quadrupole action (including counter terms), we may further consider diagonalizing the total quadrupole action including the counter terms, but there is no clear advantage of doing so, given that gravitational interactions automatically mix different sectors, and thus we leave such an analysis for the future.

\begin{figure*}[t]
    \centering
    \begin{subfigure}[t]{0.5\textwidth}
        \centering
        \includegraphics[width=\textwidth]{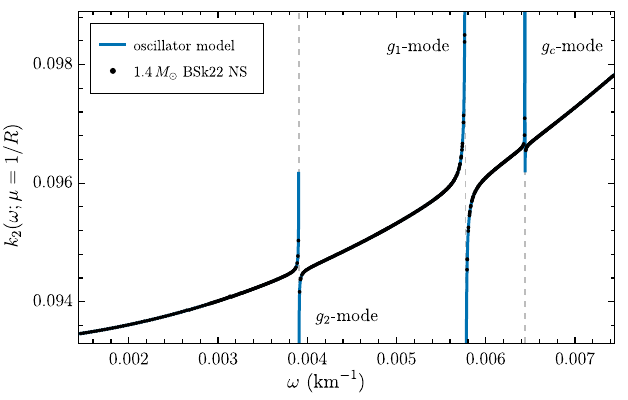}
        \caption{Low frequencies, across the $g_2$, $g_1$ and $g_c$ resonances}
    \end{subfigure}~
    \begin{subfigure}[t]{0.5\textwidth}
        \centering
        \includegraphics[width=\textwidth]{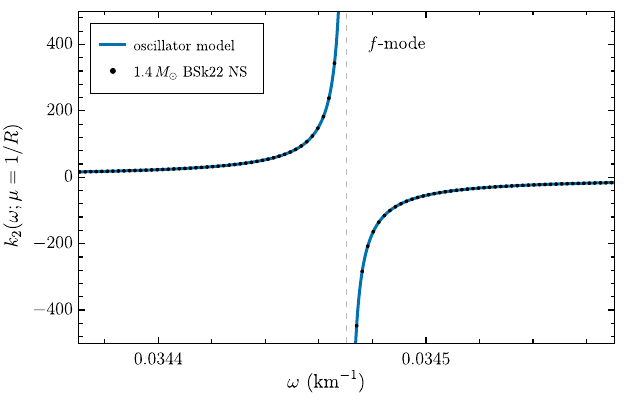}
        \caption{Close to the $f$-mode resonance}
        \label{fig:osc_f}
    \end{subfigure}
    \caption{The tidal response of the $1.4\,M_\odot$ BSk22 star near the $g_2$, $g_1$ and $g_c$ modes (a) and near the $f$-mode (b) at $\mu=R^{-1}$. The points show $k_2^{\rm generic}(\omega;\mu)$, obtained using Eq.~\eqref{eq:tidmain} with $\kappa(\mu)$ and $\sigma(\mu)$ kept in $\Sigma_{\rm R}$, and the line shows $k_2^{\rm osc.}(\omega;\mu)=\frac{3}{4}F_{\rm osc}(\omega;\mu)$ with the parameters of Table~\ref{tab:osc}. The dashed lines mark the mode frequencies and the residual is shown in Fig.~\ref{fig:osc_res}.
    }
    \label{fig:osc}
\end{figure*}

The 2PM and 4PM divergences are more subtle. On one hand, we need not add any counterterms for them in the oscillator action. We can simply replace the coupling function $g_{\rm B}(\omega)$ with that in Eq.~\eqref{eq:gsub4}.
However, unlike before, it is not trivial to absorb the freedom in $\kappa(\mu)$ and $\sigma(\mu)$ into $F_{\rm osc}(\omega;\mu)$ here because we have a particular UV model. More precisely, we can eliminate $\kappa(\mu)$ and $\sigma(\mu)$ via the rescaling $\tilde{Q}^{\mu\nu}(\omega)\rightarrow \big[1-\bar{\omega}^2\kappa(\mu)+\bar{\omega}^4\sigma(\mu)\big]^{-1}\tilde{Q}^{\mu\nu}(\omega;\mu)$, but then the quadrupole action would no longer be that of a collection of simple harmonic oscillators. Instead, it would be modified to
\begin{equation}
   \int \frac{\diff\omega}{2\pi } \sum_{n}\frac{\left[\omega^2-\omega_n^2\right]\tilde{Q}^n_{\mu\nu}(\omega)\tilde{Q}^{\mu\nu}_n(-\omega)}{4\lambda_n \omega_n^2\big[1-\bar{\omega}^2\kappa(\mu)+\bar{\omega}^4\sigma(\mu)\big]^2},
\end{equation}
This is undesirable because the frequency-dependent prefactor introduces higher-derivative terms into the action and obscures its interpretation in terms of simple harmonic modes. Thus, it is better to leave $\kappa(\mu)$ and $\sigma(\mu)$ as part of the coupling, following Ref.~\cite{Mandal:2023hqa}. This choice does not affect their relation to any observable.

Thus, we finally get the gravitationally-dressed response as identical to Eq.~\eqref{eq:QQdressed}, except with the identification in Eq.~\eqref{eq:1PMoscmap}.
We can now directly match to the scattering phase in Eq.~\eqref{eq:sfin} to fit the parameters in the oscillator model, namely $\lambda_n$, and $\omega_n$, as well as $\alpha(\mu)$, $\kappa(\mu)$, $\sigma(\mu)$ and $\alpha^{(3)}(\mu)$.

We take each $\omega_n$ from the tidal phase itself, at the frequency of the corresponding resonance, where the phase varies very rapidly. For the $g$-modes, this also fixes the response at each resonance, since the shift of the resonance due to gravitational dressing is suppressed by $(R\omega_n)^5\bar\omega_n$, and is negligible. For the $f$-mode, it fixes the response away from the resonance, while the resonance itself requires the counterterm $\alpha$, as discussed below. We then fit the $g$-mode amplitudes $\lambda_n$ to the behaviour near each $g$-mode, and fix $\lambda_f$ through the static Love number, $k_2(0)=\frac{3}{2R^5}\sum_n\lambda_n$. Next, we fit $\kappa(R^{-1})$ and $\sigma(R^{-1})$ jointly to the phase in Eq.~\eqref{eq:sfin}, with $\Sigma_{\rm R}$ including the 3PM and 4PM terms in Eqs.~\eqref{eq:SigmaR3PM} and \eqref{eq:SigmaR4PM}, in a low-frequency window above the $g$-mode frequencies, where the profile is smooth. Since $\sigma$ enters only at $\bar\omega^4$, the window extends up to $M\omega\simeq0.045$.

Near the $f$-mode, the width of the resonance follows from the imaginary part of $\Sigma_{\rm R}$, as demonstrated in Ref.~\cite{Saketh:2026trm}, while its real part shifts the resonance away from $\omega_f$ by an amount that depends on $\mu$. The counterterm $\alpha(\mu)$ absorbs this shift, and we fix $\alpha(R^{-1})$ by requiring that the real part of $F_{\rm osc}^{-1}(\omega;\mu)-\frac{1}{5}(R\omega)^5\Sigma_{\rm R}(\omega;\mu)$ vanish at $\omega=\omega_f$. Without $\alpha$, the resonance would be shifted by about $5\times10^{-6}\,\omega_f$ and $2\times10^{-5}\,\omega_f$ for the two stars, roughly $1\%$ and $2\%$ of its width, while away from the resonance $\alpha$ is irrelevant since it multiplies $\bar\omega(R\omega)^5$. The condition above fixes only the combination $\alpha(R^{-1})+\alpha^{(3)}(R^{-1})\,\bar\omega_f^2$, with $\bar\omega_f=M\omega_f$, since both couplings multiply $\bar\omega(R\omega)^5$ in Eq.~\eqref{eq:1PMoscmap} and $\bar\omega^2$ is nearly constant across the narrow resonance. We therefore set $\alpha^{(3)}(R^{-1})=0$ \footnote{Any other value is absorbed by $\alpha(R^{-1})$, up to a change in the response away from the $f$-mode of less than $10^{-2}\%$, even for $|\alpha^{(3)}(R^{-1})|=100$.}. At other scales, $\alpha^{(3)}(\mu)$ follows from Eq.~\eqref{eq:1PM3PMrun}. Close to the $f$-mode, the 4PM contributions at $\bar\omega^4$ also matter, and without them the fit deviates there by up to $0.4\%$ for the $1.4\,M_\odot$ star. With them, the deviation drops to about $0.003\%$.

\begin{table}[b]
\centering
\setlength{\tabcolsep}{7pt}      \renewcommand{\arraystretch}{1.5}

\begin{tabular}[c]{|@{\hspace{2pt}}c@{\hspace{2pt}}|@{\hspace{3pt}}c@{\hspace{3pt}}|c|c|}
\hline
\multicolumn{2}{|c|}{} & $1.4\,M_\odot$ & $1.98\,M_\odot$ \\
\hline
\multirow{3}{*}{\scriptsize\renewcommand{\arraystretch}{1}\begin{tabular}{@{}c@{}}Post-\\adiabatic\\tidal\\numbers\end{tabular}} & $\alpha$ & $-0.062208(4)$ & $0.157365(9)$ \\
\cline{2-4}
 & $\kappa$ & $1.493(2)$ & $-1.20(1)$ \\
\cline{2-4}
 & $\sigma$ & $15(3)$ & $47(9)$ \\
\hline
\multirow{4}{*}{\scriptsize\renewcommand{\arraystretch}{1}\begin{tabular}{@{}c@{}}Mode\\frequency\end{tabular}} & $\omega_f$ & $0.03447033761(1)$ & $0.04034437006(5)$ \\
\cline{2-4}
 & $\omega_{g_1}$ & $0.00577308945(6)$ & $0.007574464538(5)$ \\
\cline{2-4}
 & $\omega_{g_2}$ & $0.0039057097(2)$ & $0.0048198665(1)$ \\
\cline{2-4}
 & $\omega_{g_{c/3}}$ & $0.00643829(2)$ & $0.0036352563(7)$ \\
\hline
\multirow{4}{*}{\scriptsize\renewcommand{\arraystretch}{1}\begin{tabular}{@{}c@{}}Mode\\coupling\end{tabular}} & $\lambda_f$ & $23729.1195(7)$ & $10610.856273(8)$ \\
\cline{2-4}
 & $\lambda_{g_1}$ & $2.18507(3)$ & $0.838194(1)$ \\
\cline{2-4}
 & $\lambda_{g_2}$ & $0.36797(2)$ & $0.160024(3)$ \\
\cline{2-4}
 &  $\lambda_{g_{c/3}}$ & $0.0594(7)$ & $0.032578(7)$ \\
\hline
\end{tabular}
\caption{Parameters of the oscillator model for the $1.4\,M_\odot$ and $1.98\,M_\odot$ BSk22 stars at $\mu=R^{-1}$. The mode frequencies are the resonances in the tidal phase, $\lambda_f$ follows from the static Love number, the $g$-mode parameters come from fits near each resonance, and $\kappa(R^{-1})$ and $\sigma(R^{-1})$ come from a joint fit in the window $0.007\le\omega\le0.022$ km$^{-1}$ for the $1.4\,M_\odot$ star and $0.008\le\omega\le0.0155$ km$^{-1}$ for the $1.98\,M_\odot$ star. The rows of $g_{c/3}$ give $g_c$ for the $1.4\,M_\odot$ star and $g_3$ for the $1.98\,M_\odot$ star.
The figures in parentheses after the values describe uncertainties in the last digits.
}
\label{tab:osc}
\end{table}

\begin{figure}[t]
\centering
\includegraphics[width=\columnwidth]{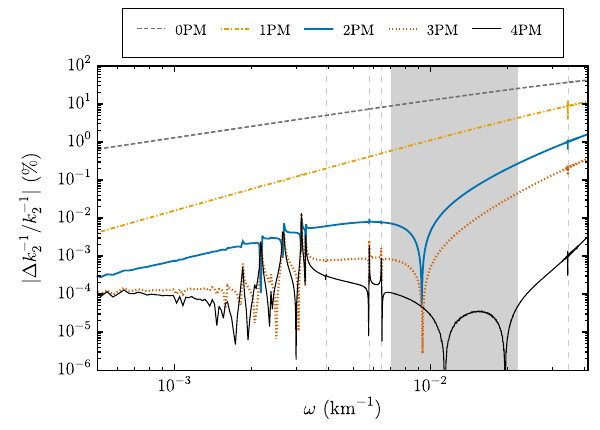}
\caption{Relative difference of the inverse responses for $1.4\,M_\odot$ BSk22 star at $\mu=R^{-1}$ over the full frequency range, $|\Delta k_2^{-1}/k_2^{-1}|$ with $\Delta k_2^{-1}/k_2^{-1}=[(k_2^{\rm generic})^{-1}-(k_2^{\rm osc.})^{-1}]\,k_2^{\rm osc.}$, in percent, with $\Sigma_{\rm R}$ truncated at 0PM to 4PM, each with its own fit. The dashed lines mark the modes and the shaded band is the window used for $\kappa(R^{-1})$ and $\sigma(R^{-1})$.}
\label{fig:osc_res}
\end{figure}

For the $1.4\,M_\odot$ and $1.98\,M_\odot$ BSk22 stars, with radii $R=13.07$ and $12.61$ km, respectively, we obtain the values in Table~\ref{tab:osc}. The errors on $\omega_n$ and $\lambda_n$ are those of the local fits near each resonance, including their change when the window of the fit is varied, and the error on $\lambda_f$ follows from those of the $g$-mode amplitudes through the static Love number. The standard errors of the joint fit of $\kappa(R^{-1})$ and $\sigma(R^{-1})$, below $10^{-4}$ and $0.1$ for the $1.4\,M_\odot$ star and $10^{-3}$ and $0.5$ for the $1.98\,M_\odot$ star, severely underestimate their uncertainty, since the phase is computed to high precision and the residuals of the fit reflect the truncation of $\Sigma_{\rm R}$ and the range of the fit window. We therefore move the upper edge of the window of the fit, shaded in Fig.~\ref{fig:osc_res}, by a third either way, and quote in Table~\ref{tab:osc} the largest change of $\kappa(R^{-1})$ and $\sigma(R^{-1})$\footnote{The sign of $\kappa$ depends on the scale. By Eq.~\eqref{eq:BetaFunckappa}, $\kappa(\mu)=\kappa(R^{-1})+\frac{214}{105}\ln(\mu R)$, so that at $\mu=r_s^{-1}$ both values are positive, $3.8$ and $0.37$. It would be interesting to better understand its dependence on neutron-star parameters in future work.}. The error on $\alpha(R^{-1})$ follows from that on $\omega_f$, since $\alpha$ shifts the resonance by $M\lambda_f\omega_f^7$ per unit $\alpha$, or $2.8\times10^{-6}$ and $5.4\times10^{-6}$ km$^{-1}$ for the two stars. Nevertheless, these errors should not be taken at face value, since they assume that the numerically computed phase is exact to machine precision and that the equation of state is known exactly\footnote{Ideally, these errors should be supplemented by the uncertainty in the interpolation of the equation of state, as well as the accuracy of the solver for the stellar perturbation equations. In this simplified exercise, we take the computed phase to be exact to the precision of the numerical computation.}. Neither is the case.

Plots showing the renormalized tidal response for the oscillator model, $k_2^{\rm osc}(\omega;\mu)=\frac{3}{4}F_{\rm osc}(\omega;\mu)$, and for the generic model, $k_2^{\rm generic}(\omega;\mu)$, are shown in Fig.~\ref{fig:osc}. Here, by $k_2^{\rm generic}$, we mean the expression obtained by inverting Eq.~\eqref{eq:sfin} with $\kappa(\mu)$ and $\sigma(\mu)$ kept in $\Sigma_{\rm R}$. In both cases, $\kappa(\mu)$ and $\sigma(\mu)$ are the values obtained, as explained before, from the fit in the low-frequency window just above the $g$-mode frequencies.
Fig.~\ref{fig:osc} shows the overlap for $\mu=R^{-1}$, but we checked it is equally good for $\mu=r_s^{-1}$. This  demonstrates consistency, since the fit was done only at $\mu=R^{-1}$ and the couplings $\alpha$, $\alpha^{(3)}$, $\kappa$ and $\sigma$ at $\mu=r_s^{-1}$ were obtained from the fitted values using the flow equations in Eqs.~\eqref{eq:BetaFunckappa},\eqref{eq:sigmaflow}, \eqref{eq:1PM3PMrun}.

Finally, Fig.~\ref{fig:osc_res} shows the residual of the fit over the full frequency range. The error remains very small, below $0.015\%$ for $\omega\ge4\times10^{-4}$ km$^{-1}$, and the systematic increase towards the $f$-mode that appears without the 4PM contributions is absent. Below this frequency, the stellar perturbation data themselves scatter at the $0.1\%$ level, and the residual reflects this numerical noise rather than the model. To show the contribution of each PM order, Fig.~\ref{fig:osc_res} also gives the residual with $\Sigma_{\rm R}$ truncated at 0PM, 1PM, 2PM and 3PM, with the same truncation in $k_2^{\rm generic}$. At each order, we keep the $\omega_n$ fixed to their values in Table~\ref{tab:osc}, refit the $g$-mode amplitudes, fix $\lambda_f$ through the static Love number, and fit only the couplings that $\Sigma_{\rm R}$ contains at that order. These are $\kappa(R^{-1})$ from 2PM, fitted alone for $\omega\le0.011$ km$^{-1}$ at 2PM and 3PM, and $\sigma(R^{-1})$ at 4PM, while $\alpha(R^{-1})$ is fixed as above from 1PM on. Note the reduction in error at larger frequencies as higher PM orders are included.

\section{Incorporation into 3PN dynamics}
\label{sec-binary}

In this section we combine the matched tidal response with the dynamics of a binary system computed up to 3PN.
We use the 3PN Hamiltonian for dynamical gravitoelectric tides from Ref.~\cite{Mandal:2023hqa}, where a single relativistic oscillator was coupled to the external tidal field via a coupling $g(\omega) Q_{\mu\nu}E^{\mu\nu}$. At 3PN, they included the $\omega^2$ term, which involves the renormalized post-adiabatic Love number $\kappa$. We take $g(\omega)E_{ij}$ from the point-particle and quadrupole-coupling sectors of the Hamiltonian in Ref.~\cite{Mandal:2023hqa}, and replace their single oscillator with the total response $F_{\rm osc}$ of Sec.~\ref{sec-osc}. We first provide the tidal field for circular orbits in the rest frame of the star. Then we use the result to compute the part of the binding energy that depends explicitly on $Q$. Finally, we take the adiabatic limit of the oscillator model and compute the full tidal binding energy at 3PN.

\subsection{Electric tidal field to 3PN}

To incorporate the tidal response in the binary dynamics, we first compute the quadrupole coupling (tidal field) in the Hamiltonian ($H$) of Ref.~\cite{Mandal:2023hqa} by reading off $g(\omega)E_{ij}=+2/z_{(1)} \,\partial H/\partial Q^{ij}_{(1)}$.
Then this tidal field, at the location of body 1 due to the companion in the binary, can be decomposed as
\begin{align}
    g~E_{ij}=\sum_{m=-2}^{2}E_m\,\mathcal{Y}^{2m}_{ij}
\end{align}
where we use the STF tensors $\mathcal{Y}^{2m}_{ij}$ defined by
\begin{equation}
  \mathcal{Y}^{2m}_{ij}\,\hat n^i\hat n^j=Y_{2m}(\theta,\phi) \ .
  \label{eq:decomp}
\end{equation}
Here, we have used the orthonormal triad $\bm n=\bm r/r$, $\bm\ell=\bm L/|\bm L|$ and $\bm\lambda=\bm\ell\times\bm n$, and measure the polar angle $\theta$ from $\bm\ell$ and the azimuth $\phi$ from $\bm n$.
In the center-of-mass frame, this tensor is built from $n^in^j$, $\ell^i\ell^j$ and $n^{(i}\ell^{j)}$, with coefficients that depend on $r$, $p_r$ and $L$. We evaluate it on circular orbits of the point-particle Hamiltonian: $p_r=0$, with $r(x)$ and $L(x)$ obtained from $\partial H_{\rm pp}/\partial r=0$ and $\Omega=\partial H_{\rm pp}/\partial L$ to 3PN, where $x=(G M_T \Omega)^{2/3}$, $\Omega$ is the orbital frequency and $M_T=m_1+m_2$. In this we neglect the tidal corrections to the orbit. The resulting mode amplitudes of the tidal field on body~1, for a companion on a circular orbit, are \cite{Mandal:2023hqa}
\begin{widetext}
\begin{align}
 \widetilde{E}_{\pm2}=&\sqrt{\frac{2\pi}{15}}\;X_2\,x^3\,\Bigg\{-3 + \left(-\frac{9}{2} + \frac{9}{2}X_2 - 3X_2^2\right)x + \left(-\frac{1403}{56} + \frac{1385}{56}X_2 - \frac{327}{28}X_2^2 + 4X_2^3 - X_2^4\right)x^2\nonumber \\
&\qquad+ x^3\Bigg[-\frac{228993}{19600} + \left(\frac{7915283}{58800} - \frac{3495\pi^2}{1024}\right)X_2 + \left(-\frac{6227777}{29400} + \frac{3495\pi^2}{1024}\right)X_2^2  \nonumber\\
&\qquad + \frac{14089}{168}X_2^3 - \frac{643}{28}X_2^4 + X_2^5 + \left(12 - 24X_2 + 12X_2^2\right) \left(\kappa_1 + \frac{214}{105}\log \frac{R x}{M_T} \right)\Bigg] + \mathcal{O}(x^4)\Bigg\}\\
\widetilde{E}_{\pm1}=&0 \\[8pt]
\widetilde{E}_0=&\sqrt{\frac{4\pi}{5}}\;X_2\,x^3\,\Bigg\{1 + \left(\frac{7}{2} - \frac{3}{2}X_2 + X_2^2\right)x + \left(\frac{561}{56} - \frac{1441}{168}X_2 + \frac{719}{84}X_2^2 - \frac{4}{3}X_2^3 + \frac{1}{3}X_2^4\right)x^2 \nonumber \\
&\qquad+ \Bigg[\frac{3049}{112} + \left(-\frac{28075}{336} + \frac{1249\pi^2}{1024}\right)X_2 + \left(\frac{16309}{168} - \frac{1249\pi^2}{1024}\right)X_2^2  - \frac{1441}{56}X_2^3 + \frac{345}{28}X_2^4 - \frac{1}{3}X_2^5\Bigg]x^3 + \mathcal{O}(x^4) \Bigg\}
\end{align}
\end{widetext}
where $\widetilde{E}_m= G^2 M_T^2 E_m$, $c=1$, and $X_a=m_a/M_T$. The $m=\pm1$ components vanish because the field is symmetric under reflection through the orbital plane. The coupling $\kappa_1$ of body~1 enters only $\widetilde E_{\pm2}$, at 3PN. The $m=0$ component is static in the orbital frame, so the $\bar\omega^2$ term of $g(\omega)$ does not contribute to it.

The modes above are given in the orbital frame. The quadrupole, however, evolves in the body frame (BF) of the star, the local nonrotating rest frame that is parallel transported along its worldline. Relative to the body frame, the orbital frame rotates about $\bm\ell$ by the angle $\bar\Phi(\tau)$.
The tidal field in the inertial frame is given by a rotation $\phi\rightarrow \phi-\bar\Phi$ which leads to $Y_{2m}(\theta,\phi-\bar\Phi)=Y_{2m}(\theta,\phi)\,e^{-im\bar\Phi}$. Hence, the effect of this transformation can be described by a phase,
\begin{align}
  E^{\rm BF}_m(\tau)&=E_m\,e^{-im\bar\Phi(\tau)}
\end{align}
Now the apparent frequency of the binary in the frame of body 1 is enhanced due to redshift ($z_1$) and decreased due to the relativistic frame dragging ($\Omega_1$) which gives an effective frequency \cite{Steinhoff:2021dsn},
\begin{align}
  \frac{\diff\bar\Phi}{\diff\tau}=\bar\Omega=&\frac{\Omega-\Omega_1}{z_1}
\end{align}
where $z_1=\diff\tau/\diff t$ is the redshift of star 1 which is computed using $z_1=\partial H/\partial m_{(1)}$ evaluated on circular orbits, whereas, the frequency $\Omega_1$ is the precession of the body frame relative to the asymptotic frame, caused by frame dragging computed using terms like $\Omega_1\,\bm S_{Q(1)}\cdot\bm\ell$ in the total Hamiltonian ($H$) given in Ref.~\cite{Mandal:2023hqa}.
The modified frequency ($\bar\Omega$) up to 3PN is given as,
\begin{align}
GM_T\bar\Omega=&x^{3/2} \bigg[1+\left(-X_2 + X_2^2\right)x \nonumber\\
&+ \left(-\frac{3}{2}X_2 + \frac{11}{6}X_2^2 - \frac{2}{3}X_2^3 + \frac{1}{3}X_2^4\right)x^{2} \nonumber\\
&+ \left(-\frac{27}{8}X_2 + \frac{99}{8}X_2^2 - 14X_2^3 + 5X_2^4\right)x^{3} \bigg] \nonumber\\
&+ \mathcal{O}(x^4)
\end{align}
For a fixed circular orbit, $\bar\Omega$ is constant and
$\bar\Phi(\tau)=\bar\Omega\,\tau$.
Hence, at fixed $x$, the star sees a slightly lower driving frequency, and each mode crossing $2\bar\Omega=\omega_n$ occurs at a slightly larger $x$ than with $\Omega$ alone.

\subsection{Contribution of the tidal quadrupole moment to 3PN in a binary system}
\label{sec:QEbin}

For a preliminary quantification of the impact of the tidal quadrupole, we need a suitable quantity that has fixed normalization and is $\mu$ independent while also being easily computable. The quadrupole moment $Q$ itself is not suitable, because it can be redefined to absorb any finite part of the coupling $g_{\rm B}(\omega)$. Instead, a suitable object is given by $(1/2)Q^{\mu\nu}g_{\rm B}(\omega)E_{\mu\nu}$, which is finite and scheme independent. This can be motivated as the part of the binding energy that couples the quadrupole moment to the tidal field, \begin{equation}
\begin{gathered}
    \mathcal{E}_Q\equiv\frac12\,Q^{ij}\,\big(gE\big)_{ij}\ .
\end{gathered}
    \label{eq:EQdef}
\end{equation}
Here, the Latin indices refer to coordinates in a comoving, locally inertial frame of the tidally deformed compact object, and $(gE)$ is the finite product $\lim g_{\rm B}E$ given to 3PN in the previous subsection. The coupling enters the Hamiltonian with the opposite sign, so $\mathcal{E}_Q$ is the part of the binding energy that couples $Q$ to the tidal field. The change of the point-particle sector caused by the quadrupole is not included in it. It can also be related to the on-shell value of the quadrupole action together with the coupling term,
\begin{equation}
    \Big(\mathcal{S}_Q+\mathcal{S}_{\rm int}\Big)\Big|_{\rm on\ shell}=\frac12\,\mathcal{S}_{\rm int}\Big|_{\rm on\ shell}=-\frac12\int\diff\tau\,\mathcal{E}_Q\ .
    \label{eq:EQonshell}
\end{equation}
The quadrupole moment in turn is computed using the oscillator action of Sec.~\ref{sec-osc}. A subtlety arises here, namely that the counter terms in the oscillator action, Eq.~\eqref{eq:bareosc3PM}, are cancelled by the divergences of the quadrupole self-energy in Eq.~\eqref{eq:dressing}. This cancellation, however, requires a consistent treatment of both the conservative and radiative sectors. This is straightforward for a single body in the Compton process, but not so simple in the case of a binary in a PN-EFT treatment, because the self-energy diagrams all become scaleless in a consistent PN expansion. Additionally, including those dressings leads to an imaginary part in the quadrupole moment, and hence in $\mathcal{E}_Q$, straining its interpretation. Thus, we simply drop the counterterms in the oscillator action, while retaining those associated with the coupling $g_{\rm B}(\omega)$.

The price associated with this is that the quantity diverges at resonances, and a more careful treatment requires incorporating radiative contributions as well \cite{Galley:2009px,Mandal:2024iug}. We leave this for future work. The quadrupole moment is then given by
\begin{equation}
\begin{gathered}
    \tilde Q_{ij}(\omega)=-\frac12\,R^5F_{\rm osc}(\omega)\,\big(g\tilde E\big)_{ij}(\omega)\ ,
\end{gathered}
    \label{eq:QbinQE}
\end{equation}
where $F_{\rm osc}(\omega)$ is the oscillator response of Eq.~\eqref{eq:Fosc} with the parameters of Table~\ref{tab:osc}, and $\kappa(R^{-1})$ from the same fit enters through $(gE)$. There are, however, additional dressings of the quadrupole due to the companion, which start at 6PN. The leading contributions arise from the diagrams
\begin{equation}
\begin{tikzpicture}[baseline=0.3cm]
  \draw[wline] (-0.15,1.0) -- (1.05,1.0);
  \draw[grav] (0,1.0) -- (0,0);
  \draw[grav] (0,0) to[out=60,in=120] (0.9,0);
  \draw[wline] (-0.15,0) -- (1.05,0);
  \node[qdot,label={[vlab]below:$Q$}] at (0,0) {};
  \node[qdot,label={[vlab]below:$Q$}] at (0.9,0) {};
  \node[mdot,label={[vlab]above:$m_2$}] at (0,1.0) {};
\end{tikzpicture}
+
\begin{tikzpicture}[baseline=0.3cm]
  \draw[wline] (-0.15,1.0) -- (1.05,1.0);
  \draw[grav] (0.9,1.0) -- (0.9,0);
  \draw[grav] (0,0) to[out=60,in=120] (0.9,0);
  \draw[wline] (-0.15,0) -- (1.05,0);
  \node[qdot,label={[vlab]below:$Q$}] at (0,0) {};
  \node[qdot,label={[vlab]below:$Q$}] at (0.9,0) {};
  \node[mdot,label={[vlab]above:$m_2$}] at (0.9,1.0) {};
\end{tikzpicture}
+
\begin{tikzpicture}[baseline=0.3cm]
  \draw[wline] (-0.15,1.0) -- (1.05,1.0);
  \draw[grav] (0.45,1.0) -- (0.45,0.45);
  \draw[grav] (0,0) to[out=75,in=180] (0.45,0.45);
  \draw[grav] (0.45,0.45) to[out=0,in=105] (0.9,0);
  \draw[wline] (-0.15,0) -- (1.05,0);
  \node[qdot,label={[vlab]below:$Q$}] at (0,0) {};
  \node[qdot,label={[vlab]below:$Q$}] at (0.9,0) {};
  \node[mdot,label={[vlab]above:$m_2$}] at (0.45,1.0) {};
\end{tikzpicture}
+
\begin{tikzpicture}[baseline=0.3cm]
  \draw[wline] (-0.15,1.0) -- (1.05,1.0);
  \draw[grav] (0,0) -- (0.45,1.0);
  \draw[grav] (0.45,1.0) -- (0.9,0);
  \draw[wline] (-0.15,0) -- (1.05,0);
  \node[qdot,label={[vlab]below:$Q$}] at (0,0) {};
  \node[qdot,label={[vlab]below:$Q$}] at (0.9,0) {};
  \node[mdot,label={[vlab]above:$m_2$}] at (0.45,1.0) {};
\end{tikzpicture}
\label{eq:m2LOQE}
\end{equation}
where the upper dotted line is the worldline of the companion. This is a very small effect, and it does not remove the divergence at the resonant frequencies, so we neglect it in what follows.

To evaluate $\mathcal{E}_Q$ along a circular orbit, we decompose both $Q_{ij}$ and $(gE)_{ij}$ in the basis of Eq.~\eqref{eq:decomp}. Using the orthogonality of the STF tensors, the reality conditions $E_{-m}=(-1)^mE_m^*$ and $Q_{-m}=(-1)^mQ_m^*$, and Eq.~\eqref{eq:QbinQE} mode by mode, we obtain
\begin{align}
    &\mathcal{Y}^{2m}_{ij}\mathcal{Y}^{2m'*}_{ij}=\frac{15}{8\pi}\,\delta_{mm'},\qquad k_2^{\rm osc}=\frac34F_{\rm osc},\\
    Q_m&=-\frac23R^5\,k_2^{\rm osc}(m\bar\Omega)\,E_m,\\
    \mathcal{E}_Q&=\frac{15}{16\pi}\sum_{m=-2}^{2}Q_m\,E_m^{*} =-\frac{5R^5}{8\pi}\sum_{m=-2}^{2}k_2^{\rm osc}(m\bar\Omega)\,|E_m|^2\nonumber\\
    &=-\frac{5R^5}{8\pi}\Big[k_2^{\rm s}\,|E_0|^2+2\,k_2^{\rm osc}(2\bar\Omega)\,|E_2|^2\Big]\ ,
    \label{eq:EQmsum}
\end{align}
where in the last line we used $E_{\pm1}=0$ on circular orbits and $k_2^{\rm osc}(0)=k_2^{\rm s}$. Each term pairs $Q_m\propto e^{-im\bar\Phi}$ with $E_m^*\propto e^{im\bar\Phi}$, so $\mathcal{E}_Q$ is constant along a circular orbit. The $m=0$ term involves only the static response, while the $m=\pm2$ terms carry the dynamical tide at $\omega=2\bar\Omega$.

\begin{figure*}[t]
\includegraphics[width=1.6\columnwidth]{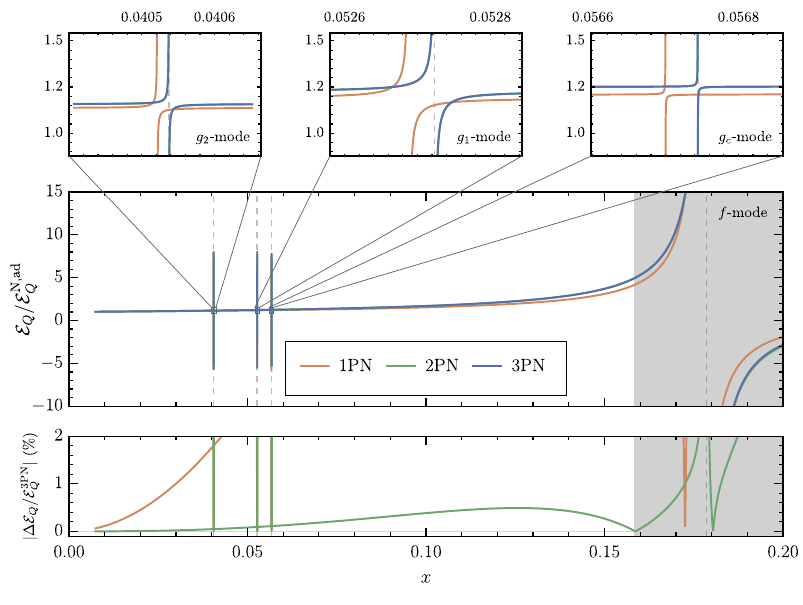}
\caption{The upper panel shows the tidal interaction energy relative to its Newtonian adiabatic value, $\mathcal{E}_Q/\mathcal{E}_Q^{\rm N,ad}$ for one star in an equal-mass binary of two $1.4\,M_\odot$ BSk22 stars on circular orbits, up to contact, at 1PN, 2PN and 3PN. The lower panel shows the relative difference from the 3PN result in percent. The insets zoom in on the $g$-mode crossings, marked by the dashed vertical lines at $2\bar\Omega=\omega_g$.
The shaded region depicts where the two stars contact, $x>x_c$, and the dashed line inside it marks the $f$-mode, $2\bar\Omega=\omega_f$.}
\label{fig:QEbin}
\end{figure*}

As before, we normalize by the Newtonian adiabatic value, obtained with $k_2^{\rm s}$ and the leading-order field, for which $|E_0|^2+2|E_2|^2=\frac{16\pi}{5}X_2^2x^6/M_T^4$. This gives
\begin{align}
    \mathcal{E}_Q^{\rm N,ad}&=-2\,k_2^{\rm s}\,R^5\,\frac{X_2^2\,x^6}{M_T^4},\\
    \frac{\mathcal{E}_Q}{\mathcal{E}_Q^{\rm N,ad}}&=\frac{5}{16\pi\,X_2^2\,x^6}\bigg[\big|\widetilde E_0\big|^2+2\,\frac{k_2^{\rm osc}(2\bar\Omega)}{k_2^{\rm s}}\,\big|\widetilde E_2\big|^2\bigg].
    \label{eq:EQratio}
\end{align}
We consider equal-mass binaries of two $1.4\,M_\odot$ and of two $1.98\,M_\odot$ BSk22 stars on circular orbits, up to contact, which we estimate at Newtonian order as $x_c=M_T/(2R)$. We evaluate Eq.~\eqref{eq:EQratio} at 1PN, 2PN and 3PN, keeping the tidal field and $\bar\Omega(x)$ to the same order, with $\kappa_1=\kappa(R^{-1})$ from the fit in the 3PN field. The fitted $\sigma(R^{-1})$ is not included, since it would enter $(gE)$ only at relative order $\bar\omega^4$, that is 6PN, beyond the order considered here. Since the $g$-mode poles of $k_2^{\rm osc}$ are real, the curves are cut in a small neighbourhood of the crossings $2\bar\Omega=\omega_g$, which are marked by dashed lines.
The results for the $1.4\,M_\odot$ binary are shown in Fig.~\ref{fig:QEbin}, and those for the $1.98\,M_\odot$ binary can be computed similarly but are not shown here.

These results show that the post-Newtonian corrections to the tidal field remain significant up to 3PN. Near contact, the 1PN result differs from the 3PN result by as much as 16\% for the $1.4\,M_\odot$ binary and 30\% for the $1.98\,M_\odot$ binary. The deviation therefore grows with the mass of the stars, as expected, since more massive stars are more compact, and thus come into contact at larger $x$. For nonspinning stars, however, the driving frequency $2\Omega$ stays below the $f$-mode all the way to contact. These plots would be far more interesting for spinning stars with antialigned spins, where the $f$-mode resonance can occur during the inspiral~\cite{Steinhoff:2021dsn,Yu:2024uxt,Markin:2026sdm}. The radiative contributions neglected here are then likely to make a significant difference, since they set the width of the resonance.

\subsection{Post-adiabatic expansion and binding energy at 3PN}
\label{sec:Ebin}

While $\mathcal{E}_Q$ presented in the previous subsection is scheme independent, it is not yet an observable in the strict sense. In particular, it does not capture the total tidal contribution to the binding energy of the orbit, since the tidal effects on the orbital parameters are not taken into account. The details are currently not known in full generality. However, in the adiabatic limit, the complete binding energy is known to 3PN~\cite{Mandal:2023hqa}. In order to make a comparison, we therefore take the adiabatic limit of the oscillator model itself.

Integrating out $Q$ with Eq.~\eqref{eq:QbinQE} in the on-shell action of Eq.~\eqref{eq:EQonshell}, without the oscillator counter terms as in Sec.~\ref{sec:QEbin}, gives an action for the tidal field alone,
\begin{equation}
\mathcal{S}_{\rm eff}[E]=\frac{R^5}{8}\int\frac{\diff\omega}{2\pi}\,F_{\rm osc}(\omega)\,g(\omega)g(-\omega)\,\tilde E_{ij}(\omega)\,\tilde E^{ij}(-\omega)\ .
\label{eq:Seff}
\end{equation}
Expanding to second order in frequency, with Eq.~\eqref{eq:Fosc} and $g(\omega)=1-\bar\omega^2\kappa(R^{-1})+\mathcal{O}(\bar\omega^4)$, gives the adiabatic action,
\begin{align}
    \mathcal{S}_{\rm eff}&=\int\diff\tau\,\frac{\lambda}{4}\Big[E_{ij}E^{ij}-2M^2\kappa_{\rm eff}\,\dot E_{ij}\dot E^{ij}\Big]+\dots
\end{align}
where,
\begin{equation}
\begin{split}
\lambda&=\sum_n\lambda_n=\frac23k_2^{\rm s}R^5,\qquad \zeta=\frac{\sum_n\lambda_n\omega_n^{-2}}{\sum_n\lambda_n},\\
\kappa_{\rm eff}&=\kappa(R^{-1})-\frac{\zeta}{2M^2}\ .
\end{split}
\label{eq:Seffad}
\end{equation}
Here $M$ is the mass of the star, so $\bar\omega=M\omega$, and the dots stand for terms with four or more time derivatives, which enter at relative 6PN. The $\omega^2$ term of the oscillator response does not depend on $\mu$, so $\kappa_{\rm eff}$ runs as $\kappa$ in Eq.~\eqref{eq:BetaFunckappa}. Both terms in $\kappa_{\rm eff}$ multiply the same operator, so the binary cannot separate them. We obtain $\zeta/(2M^2)=98.9$ and $36.1$, and $\kappa_{\rm eff}(R^{-1})=-97.4$ and $-37.3$, for the $1.4\,M_\odot$ and $1.98\,M_\odot$ stars. It is worth noting that the $f$-mode is responsible for more than $99\%$ of $\zeta$, consistent with the dominance of the $f$-mode in relativistic mode sums~\cite{HegadeKR:2026kku} and overlap integrals~\cite{Andersson:2019ahb}, and with Ref.~\cite{PitrePoisson2024}, where the effective frequency defined from the static and post-adiabatic Love numbers lies close to the $f$-mode frequency, and is related to the static deformability in an approximately equation-of-state independent way~\cite{Saes:2025jvr}.

Mapping to the static and dynamical Love numbers of Ref.~\cite{Apostolidis:2026qsg}, defined by the action $\frac12\int\diff\tau\,\big[R^5c_EE_{ij}E^{ij}+(R^8/M)\,c_{\dot E}\dot E_{ij}\dot E^{ij}\big]$, we obtain $c_E=k_2^{\rm s}/3$ and, approximately, $c_{\dot E}\simeq(M/R)^3c_E\big(\zeta/M^2-2\kappa(R^{-1})\big)$ at $\mu=R^{-1}$. For the $1.4\,M_\odot$ and $1.98\,M_\odot$ stars, we find $c_E=0.0311$ and $0.0167$, and $c_{\dot E}\simeq0.0240$ and $0.0155$.

Mapping to the conventions used in the analysis of neutron-star scattering in Ref.~\cite{Schulze:2026ewu}, which uses the same subtraction scheme, we obtain $\mu^{(2)}=\lambda$ and $\kappa_{\dot E^2}(R_0)\equiv\mu'^{(2)}/(M^2\mu^{(2)})=-2\kappa_{\rm eff}(R_0^{-1})$. They estimated $\kappa_{\dot E^2}(R)\simeq1/(M\omega_f)^2\sim+100$ from a normal-mode picture, which neglects the logarithmic running, and found that such values increase the tidal scattering angle and worsen the agreement with numerical relativity. Our matched values at $\mu=R^{-1}$, $194.8$ and $74.6$, lie within about $1\%$ and $4\%$ of that estimate, $196.9$ and $71.9$, since $\kappa(R^{-1})$ is small. The normal-mode estimate therefore works well at this scale, but it cannot hold at all scales, since $\kappa_{\dot E^2}$ runs by $\frac{428}{105}\ln(R_0/R)$, for example by $-4.7$ and $-3.1$ at $R_0=r_s$. Our values come with this running, which they noted is missing from the post-adiabatic Love numbers available so far. A quantitative comparison with the post-adiabatic Love numbers of Refs.~\cite{PitrePoisson2024,Saes:2025jvr} requires mapping conventions, finding common quantities, and using common equation-of-state(s), which we leave for future work.

Eq.~\eqref{eq:Seffad} has the form of the adiabatic tidal action of Ref.~\cite{Mandal:2023hqa}, so their binding energy for circular orbits applies with their Love numbers replaced by $\hat\lambda_1=\lambda/M_T^5$ and $\kappa_{\rm eff}(R^{-1})$. Its tidal part for the first star, per reduced mass $\nu M_T$ with $\nu=X_1X_2$, is
\begin{widetext}
\begin{align}
\frac{E_{\rm tid}^{(1)}}{\nu M_T}=& \frac{\hat\lambda_1 x^6}{1-X_2}\Bigg\{
\frac{9}{2} X_2
+ \left(22X_2 - 22X_2^2 + \frac{33}{4}X_2^3\right) x + \left(\frac{1014}{7}X_2 - \frac{30745}{168}X_2^2 + \frac{19331}{168}X_2^3 - \frac{247}{6}X_2^4 + \frac{91}{16}X_2^5\right) x^2 \nonumber\\
&\qquad\qquad+ \Bigg[\frac{19898}{49}X_2 + \left(-\frac{580203}{392} + \frac{8895\pi^2}{512}\right) X_2^2 + \left(\frac{703539}{392} - \frac{8895\pi^2}{512}\right) X_2^3 - \frac{85405}{112}X_2^4 + \frac{26315}{112}X_2^5  \nonumber\\
&\qquad\qquad- 30X_2^6+ \frac{45}{32}X_2^7 + \left(-45X_2 + 90X_2^2 - 45X_2^3\right) \left(\kappa_{\rm eff}(R^{-1}) + \frac{214}{105}\log \frac{Rx}{M_T} \right) \Bigg] x^3 + \mathcal{O}(x^4)
\Bigg\}.
\label{eq:Ebin}
\end{align}
\end{widetext}
Here, $E_{\rm tid}^{(2)}$ follows by exchanging the two bodies, and for the equal-mass binaries considered here $E_{\rm tid}=E_{\rm tid}^{(1)}+E_{\rm tid}^{(2)}=2E_{\rm tid}^{(1)}$. The combination $\kappa_{\rm eff}(R^{-1})+\frac{214}{105}\log(Rx/M_T)$ is $\kappa_{\rm eff}$ at the orbital scale $\mu=x/M_T$. Since $\kappa_{\rm eff}(\mu)=\kappa_{\rm eff}(R^{-1})+\frac{214}{105}\log(\mu R)$, starting from any other scale gives the same combination, $\kappa_{\rm eff}(\mu)+\frac{214}{105}\log\big(x/(\mu M_T)\big)$, so $E_{\rm tid}$ does not depend on the choice $\mu=R^{-1}$.

The 3PN term is where the frequency dependence of the response first enters the binding energy, through $\kappa_{\rm eff}$, which is dominated by the $f$-mode. The matched values $\kappa(R^{-1})=1.49$ ($-1.20$) shift $\kappa_{\rm eff}$ from the value set by $-\zeta/(2M^2)$ by about $1.5\%$ ($3\%$).

\begin{figure}[t]
\centering
\includegraphics[width=\columnwidth]{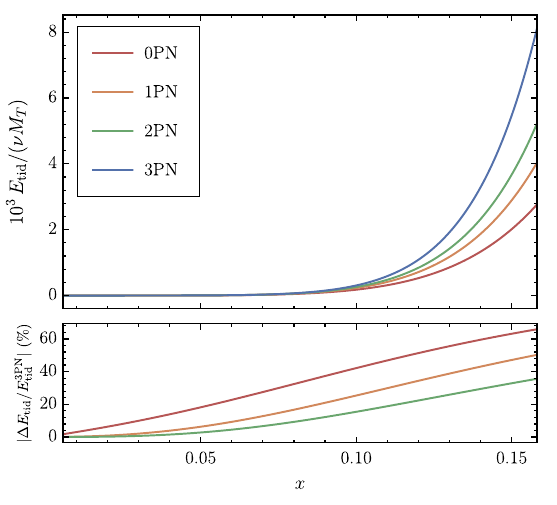}
\caption{The upper panel shows the adiabatic tidal binding energy $E_{\rm tid}/(\nu M_T)$ of Eq.~\eqref{eq:Ebin}, summed over both stars, for an equal-mass binary of two $1.4\,M_\odot$ BSk22 stars on circular orbits, from $x\simeq0$ up to contact at Newtonian order, $x_c=M_T/(2R)\simeq0.158$, or a gravitational-wave frequency of about $1.45$ kHz, at 0PN, 1PN, 2PN and 3PN, with $\kappa_{\rm eff}(R^{-1})$ of Eq.~\eqref{eq:Seffad}. The lower panel shows the relative difference from the 3PN result in percent.}
\label{fig:Ebin}
\end{figure}

We evaluate $E_{\rm tid}/(\nu M_T)$ for the two binaries of Sec.~\ref{sec:QEbin}, up to contact, at Newtonian order and at 1PN, 2PN and 3PN, with the fitted $\kappa(R^{-1})$ and mode parameters.
The results for the $1.4\,M_\odot$ binary are shown in Fig.~\ref{fig:Ebin}, and those for the $1.98\,M_\odot$ binary can be computed similarly but are not shown here.

It is clear from the figures that post-Newtonian corrections make a notable difference in the post-adiabatic binding energy, with a relative difference of about $66\%$ between the 3PN and 0PN results near contact for $1.4\,M_\odot$. It is important to keep in mind that the difference is likely even greater, owing to the low-frequency expansion used in this subsection. It is therefore desirable to compute the binding energy without such an approximation, so that the effect of the resonant oscillation modes, including the $g$-modes, is taken into account.

\section{Conclusions}
\label{sec-conclusion}
Dynamical tides probe the oscillation modes of a neutron star, and hence its internal composition, but using them in gravitational-wave models requires a relativistic tidal response that captures the resonances and can be systematically incorporated into binary dynamics. In a worldline EFT, dominant, electric, quadrupolar tidal effects are captured by a quadrupolar degree of freedom with the tidal-response function defined as its propagator.

We have obtained the dynamical tidal response of a neutron star to relative 4PM order, extending the procedure we established in Ref.~\cite{Saketh:2026trm}. There, the response is matched to the tidal scattering phase, defined as the difference between the gravitational Compton/Raman scattering phases of a neutron star and a black hole. Starting from 1PM, the EFT computation of this phase encounters divergences at every order, which renormalize either the tidal-response function or the coupling of the quadrupole to the tidal field. Since any finite modification to the coupling can be absorbed by redefining the tidal-response function, this leaves behind a single undetermined function, which we fixed in terms of the tidal scattering phase. We used this to obtain the renormalized response for a $1.4\,M_\odot$ BSk22 star , and studied how it shifts as the renormalization scale $\mu$ is varied from $\mu r_s=10^{-2}$ to $\mu r_s=1$. We chose $\mu=R^{-1}$ as a suitable scale for the main analysis, since this provides a natural length scale separating the stellar interior from the exterior.

We then showed that the tidal-response function can also be mapped to that of an explicit action of relativistic harmonic oscillators for both the $f$-mode and $g$-modes, supplemented by counterterms that absorb the gravitational dressing. The mode parameters in the oscillator model, along with $\kappa(R^{-1})$ and $\sigma(R^{-1})$ arising from the modified coupling, can be fitted numerically at $\mu=R^{-1}$ from the tidal scattering phase. The resulting oscillator model reproduces the full response to better than $0.015\%$ for $\omega\ge4\times10^{-4}$ km$^{-1}$, and tracks it at other scales through the renormalization-group flow equations.

Finally, we showed how the tidal-response function in the oscillator picture can be systematically incorporated into binary dynamics at 3PN order using the electric tidal field previously derived in Ref.~\cite{Mandal:2023hqa}. We used the result to compute the coupling energy of the induced quadrupole, $\mathcal{E}_Q$, which is proportional to the on-shell action of the quadrupole sector together with the coupling term, and is $\mu$-independent, for equal-mass quasicircular binaries of $1.4\,M_\odot$ and $1.98\,M_\odot$ stars at PN orders up to 3PN. We found that this tidal energy of the induced quadrupole changes by up to $16\%$ and $30\%$ between 1PN and 3PN near contact, for the lighter and heavier binary system, respectively. While this is a suitable way to demonstrate the impact of post-Newtonian corrections to the tidal effects, it is not in itself an observable, since it misses the contribution of the tidal effects to the orbit. Thus, separately, working in the post-adiabatic limit, we computed the total tidal contribution to the binding energy using Ref.~\cite{Mandal:2023hqa}, and showed that the 3PN and Newtonian adiabatic tidal binding energies differ by about $66\%$ ($75\%$) for the lighter (heavier) binary towards the end of the inspiral, at contact.

In the post-adiabatic limit, we also mapped our tidal-response function to previous works that studied neutron-star tides in the adiabatic regime, and showed that the matched post-adiabatic Love number agrees within about $1\%$ and $4\%$ with the normal-mode estimate set by the $f$-mode frequency, as the $f$-mode accounts for more than $99\%$ of its oscillator part.

This work can be extended in several directions. The most immediate would be the calculation of the complete tidal binding energy at 3PN without the adiabatic expansion, which requires the change of the point-particle sector caused by the quadrupole and would capture the $g$-mode contributions. Additionally, while the Compton scattering used to obtain the tidal response includes all radiative effects, the incorporation of the response into the 3PN dynamics drops these radiative contributions, since it assumes a clean separation of the conservative and dissipative sectors. This is not valid at resonant frequencies, as the width of each resonance is set by dissipation, and neglecting it leads to poles in the tidal-response function. We also need to include viscous dissipation in the stellar interior~\cite{Ghosh:2023vrx,Ghosh:2025wfx}, whose tidal response has so far been matched to the worldline EFT only at leading order in a low-frequency expansion~\cite{Saketh:2024juq,Savic:2026cgy}. It is also worth investigating how the mode amplitudes in this work, which are EFT couplings fixed by a gauge-invariant scattering observable, relate to the overlap integrals of relativistic mode-sum calculations~\cite{HegadeKR:2025qwj,HegadeKR:2026kku}, which are defined within stellar perturbation theory, since the two approaches need to be matched explicitly.

A particularly important direction involves the extension to spinning neutron stars. On the EFT side this extension could be straightforward, while a proper EFT matching with perturbation theory for spinning neutron stars has not been attempted, even for adiabatic tides. For anti-aligned spins, the $f$-mode resonance can be excited within the inspiral~\cite{Steinhoff:2021dsn,Yu:2024uxt,Markin:2026sdm}, and rotation brings in frame dragging and resonances associated with inertial modes. Equally important, and more involved, is incorporating the renormalized dynamical tides into waveform models to quantify their impact on gravitational-wave observations. Finally, the framework can be applied to a range of equations of state to understand how the various oscillator parameters vary. It would also be interesting to test the dynamical tidal response obtained here against the bounds on dynamical Love numbers derived recently from the surface impedance~\cite{McLoughlin:2026cnd}. We plan to explore some of these directions in future work.
\section*{acknowledgements}
We thank Emil Bjerrum-Bohr, Gang Chen, Thibault Damour, Alex Edison, Abhishek Hegade K.~R., Nabha Shah and Nicol\'as Yunes for useful discussions and feedback.
The research of M.V.S.S. is supported by the National Post-Doctoral Fellowship (PDF/2025/004764), ANRF, Government of India.
N.A. and S.G. gratefully acknowledge support from the STFC via Grant No. ST/Y00082X/1.
G.M. and M.V.S.S. acknowledge support from the Department of Atomic Energy, Government of India, under project no. RTI4019.
R.P.’s research is funded by, the Deutsche Forschungsgemeinschaft (DFG, German Research Foundation), Projektnummer 417533893/GRK2575 ``Rethinking Quantum Field Theory'' and the VILLUM Foundation (grant no. VIL37766); The Center of Gravity is a Center of Excellence funded by the Danish National Research Foundation under grant no. DNRF184.
The authors acknowledge the use of Claude \cite{Anthropic2026Claude} and OpenAI \cite{OpenAI2026ChatGPT} to assist in computing and optimizing code and in proofreading the manuscript. All code and text was reviewed by the authors, who take full responsibility for
the scientific content, accuracy, and integrity of this paper.

\bibliographystyle{apsrev4-2}
\bibliography{refv2}
\end{document}